\documentclass[submission, Phys]{SciPost}

\usepackage{chemformula}
\usepackage{bm}
\usepackage{caption}
\usepackage{subcaption}
\usepackage{amssymb}

\usepackage[normalem]{ulem}

\def\a{\alpha}

\def\p{\pi}

\def\vp{\varphi}

\def\ra{\rightarrow}

\def\<{\langle}
\def\>{\rangle}

\def\vecs{\mathbf{S}}

\usepackage{xcolor}

\begin{document}

\begin{center}{\Large \textbf{Thermodynamic phase diagram of the competition between superconductivity and charge order in cuprates
}}\end{center}

\begin{center}
Giulia Venditti\textsuperscript{1*},
Ilaria Maccari\textsuperscript{2},
Jose Lorenzana\textsuperscript{3$\dagger$} and
Sergio Caprara\textsuperscript{3}
\end{center}

\begin{center}
{\bf 1} SPIN-CNR Institute for Superconducting and other Innovative 
 Materials and Devices, Area della Ricerca di Tor Vergata, Via del Fosso del Cavaliere 
 100, 00133 Rome, Italy
\\
{\bf 2} Department  of  Physics,  Stockholm  University,  Stockholm  SE-10691,  Sweden
\\
{\bf 3} ISC-CNR and Department of Physics, Sapienza University of Rome, 
 Piazzale Aldo Moro 2, 00185, Rome, Italy
\\
*giulia.venditti@spin.cnr.it,\, $^\dagger$jose.lorenzana@cnr.it

\end{center}
	
\begin{center}
\today
\end{center}


\section*{Abstract}
{\bf
We argue that there is a special doping point in the phase diagram 
of cuprates, such that the condensation 
of holes into a charge-ordered and into a superconducting phase are degenerate 
in energy but with an energy barrier in between. We present Monte Carlo simulations of this problem without and with quenched disorder in two-dimensions. 
While in the clean case, charge order and superconductivity are separated by a 
first-order line which is nearly independent of temperature, in the presence of 
quenched disorder, charge order is 
fragmented into domains separated by superconducting filaments reminiscent of the supersolid behaviour in $^4$He. Assuming weak interlayer couplings, the resulting three-dimensional phase diagram is in good agreement with the experiments. }

\vspace{10pt}
\noindent\rule{\textwidth}{1pt}
\tableofcontents\thispagestyle{fancy}
\noindent\rule{\textwidth}{1pt}
\vspace{10pt}

\section{Introduction}
\label{intro}

\begin{figure}[bh]
    \centering 
    \includegraphics[width=.5\linewidth]{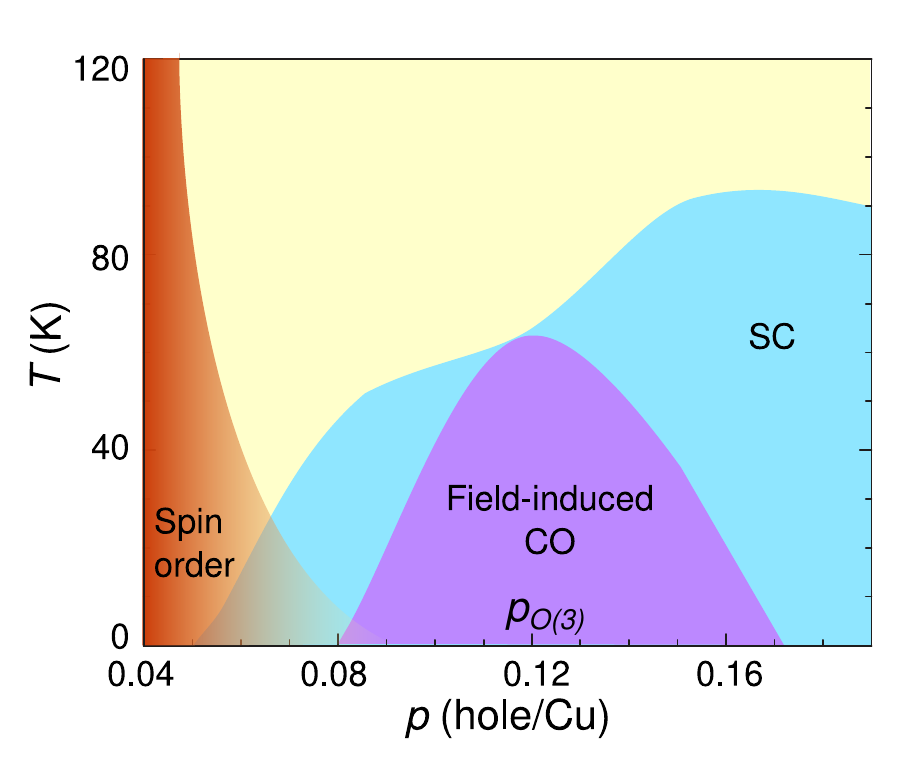} 
    \caption{a) Phase diagram of YBa$_2$Cu$_3$O$_y$ in the temperature vs. hole doping plane (the figure 
    is adapted from Ref. \cite{wu2011magnetic}).  The magenta region is the CO induced by a magnetic field so the phase diagram can be seen as a two-dimensional projection of a three-dimensional phase diagram with the magnetic field axis running perpendicular to the plane of the figure. $P_{O(3)}$ indicates the ``$O(3)$"  doping at which CO and SC are nearly degenerate.  
    }
    \label{fig:phasediag_exp_vsx}
\end{figure}

There is an overwhelming experimental 
evidence\cite{Ghiringhelli2012,chang2012direct,Leboeuf2013,Grissonnanche2014,
Forgan2015,Kacmarcik2018,caprara2020doping,Leridon2020} 
that competition between charge order (CO) and superconductivity (SC) occurs 
in high-critical-temperature superconducting cuprates. It has been 
argued\cite{attanasi2009competition,Sachdev2013,Efetov2013,caprara2020doping,
Leridon2020} that, under certain circumstances, the superconducting order 
parameter with $U(1)$ symmetry and a commensurate charge-density-wave (CDW) 
parameter with $Z_2$ (Ising) symmetry can be encoded in a single order parameter 
with approximate $O(3)$ symmetry. 
Evidence for this emergent symmetry stems from 
studies where the balance between SC and CO is controlled by a non-thermal parameter 
which couples non-linearly to one of the 
orders~\cite{Leboeuf2013,Grissonnanche2014,Gerber2015,Chang2016,Kacmarcik2018,
caprara2020doping,Leridon2020}. 
For example, a uniform magnetic field $H$ disfavours SC with respect to CO. At zero 
magnetic field, upon reducing the temperature, the 
correlation length of CO starts to grow at a temperature $T_\text{QC}$ larger than 
the superconducting critical temperature $T_\text{c}$, as if the system was 
approaching a charge-ordered state~\cite{arpaia2019dynamical}. With further lowering the temperature, {however,}
the growth {of the CO correlation-length}  stops near $T_\text{c}$, {where no CO but rather SC develops.}
Extrapolating the divergence of the CO correlation length to the superconducting 
region shows that the temperature $T_\text{CO}$ of the putative charge-ordered 
state coincides with the actual three-dimensional ordering temperature $T_{CO}^{3D}$ 
that is reached once SC is suppressed by a sufficiently strong magnetic field. One fact 
that points to an approximate $O(3)$ symmetry between the two orders is that the critical 
temperature for the field-induced CO (and therefore the putative charge-ordering 
temperature) obeys $T_{CO}^{3D}\approx T_\text{c}$ near a hole doping content, hereafter 
``the $O(3)$ point", $p_{O(3)}\approx 0.12$. 
As schematically shown in Fig.~\ref{fig:phasediag_exp_vsx}, this has been 
seen in \ch{YBa2Cu3O_{6+x}} with various probes as nuclear magnetic 
resonance~\cite{wu2011magnetic},  sound velocity, and Hall effect measurements (see 
Ref.\,\cite{Laliberte2018}, and references therein). 
The degeneracy of 
the ordering temperature at the $p_{O(3)}$ point strongly suggests that the tendency 
towards CO and SC are the same instability manifesting in different channels.

\begin{figure}
    \centering
    \includegraphics[width=0.9\linewidth]{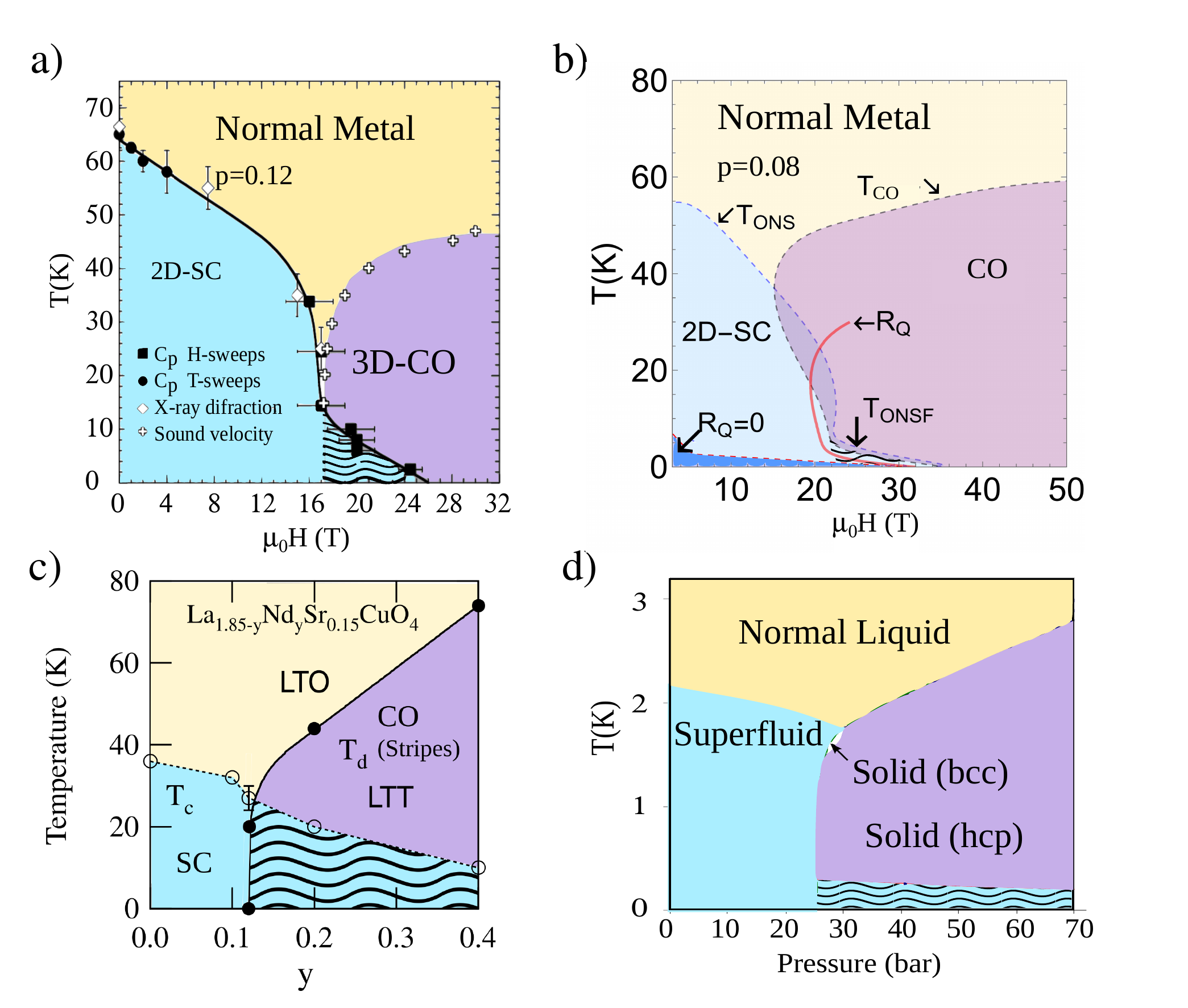} 
    \caption{ Phase diagrams showing competition between SC/superfluidity and CO/solid 
    phases. 
    a) YBa$_2$Cu$_3$O$_y$ with $y=6.67$. The onset of superconducting correlations (labelled 2D-SC) 
    was detected by an anomaly in the specific heat (adapted from 
    Ref.\,\cite{Kacmarcik2018}). Three-dimensional CO (3D-CO) was detected 
    with X-ray diffraction~\cite{chang2012direct} and  sound velocity~\cite{Leboeuf2013}.  
    b)  La$_{2-x}$Sr$_{x}$CuO$_4$ with 
 $x=0.08$. The onset temperatures for SC ($T_\mathrm{ONS}$), FSC ($T_\mathrm{ONSF}$), and CO 
 ($T_\mathrm{CO}$) were extracted from magnetoresistance data (adapted from 
 Ref.~\cite{Leridon2020}). The darker blue region corresponds to zero sheet 
 resistance $R_\square = 0$. The red line $R_\mathrm{Q}$ shows the locus of the quantum of 
 resistance.  c) La$_{1.85-y}$Nd$_y$Sr$_{0.15}$CuO$_4$. Here, isoelectronic doping ($y$) 
 favours the LTT, which stabilizes CO in the form of 
 stripes\cite{Tranquada}. 
Open circles show $T_\text{c}$ while solid circles show the structural transition 
$T_\text{d}$ which is known to be close to the CO temperature 
(adapted 
from Ref.\,\cite{Tajima2001}). 
    d) Phase diagram of $^4$He (adapted from Ref.~\cite{Kirichek2009}). In a)-c) the wavy-shaded regions denote the coexistence of CO and SC which in   Ref.~\cite{Leridon2020,caprara2020doping} was attributed to 
    filamentary superconductivity. In d) the wavy shading is the $^4$He-analogous supersolid 
    phase observed in Ref.~\cite{Kim2004b}. }
    \label{fig:phasediag_exp}
\end{figure}

It is useful to visualize this phase diagram with an extra control-parameter axis, such as the magnetic field,
perpendicular to the $T$ vs. doping plane. 
Fig.~\ref{fig:phasediag_exp} shows various cuprate experimental phase diagrams (a-c) 
and compare them with $^4$He (d). In panels (a) and (b), the control parameter 
is the magnetic field, which tunes the SC energy. We will refer to this situation as a SC-driven transition. In contrast, in panel (c) the control parameter is the isoelectronic doping, favouring 
stripes~\cite{Tranquada,Tajima2001}, which will be referred to as a CO-driven transition.
The superconducting and charge-ordered phases meet the disordered phase at a 
so-called ``bicritical" point\cite{fisher1977critical}, analogous to the bicritical 
point in $^4$He (d). An alternative terminology is that of triple point, usually adopted when the transition lines are first-order, as for instance the point 
where the liquid, gas, and solid phases of a substance meet.

{In all the phase diagrams of Fig.~\ref{fig:phasediag_exp} (a-c), a} superconducting foot develops underneath the charge-ordered state at low 
temperatures [wavy shading in (a-c)], which in 
Refs.~\cite{caprara2020doping,Leridon2020}
was associated with the occurrence of filamentary SC (FSC). This is attributed to a 
{\it tertius gaudens} (rejoicing third) effect. 
The quenched disorder breaks the CO into domains hosting different variants of CO related by discrete translations. At the interface between two variants of CO, both get frustrated and  SC is stabilized.
The same principle is believed~\cite{balibar2008supersolidity} to explain the appearance of supersolid 
phases in $^4$He [wavy shading in (d)]. 

A striking characteristic of the phase diagrams in Fig.~\ref{fig:phasediag_exp} is 
the nearly vertical separation between CO/solid and SC/superfluid phases in all the 
phase diagrams, once FSC is disregarded.  
In cuprates, several probes near $p_{O(3)}$ show that, in a wide temperature range, the critical 
magnetic field $H_\text{c}^*$, at which a long-range charge-ordered phase
stabilizes, is nearly independent of the temperature. 
This is seen in sound-velocity data \cite{Leboeuf2013}, resonant inelastic X-ray
scattering \cite{Chang2016} and nuclear magnetic 
resonance\cite{wu2011magnetic,Kacmarcik2018} experiments on 
\ch{YBa2Cu3O_{6+x}}, as shown in Fig.~\ref{fig:phasediag_exp}(a).
The transition to the superconducting phase has been determined by 
the anomaly in the density of states probed by specific heat 
measurements, which coincides with $T_\text{c}$ determined from other 
methods (see Ref.\,\cite{Kacmarcik2018}, and 
references therein).  A similar phase diagram can be deduced from  
magnetotransport experiments~\cite{shi2014two,shi2020vortex,Leridon2020,
caprara2020doping} in La-based cuprates. Here, the lines are not sharp 
anomalies, probably due to stronger disorder, but the general 
topology is the same [see Fig.\,\ref{fig:phasediag_exp}(b)]. 

{Applying a} magnetic field is not the only way to tune the balance between 
CO and SC. An alternative {path} is the structural 
enhancement of CO introduced by Tranquada and 
collaborators \cite{Tranquada1995}. In this case, isovalent doping 
induces a structural distortion which couples with CO. When plotted against the isovalent doping 
concentration \cite{Tajima2001,Fujita2002,Zhang2005,Kimura2004}, the phase
diagrams bear a striking resemblance with the magnetic-field-controlled phase 
diagrams for similar hole content [see Fig.\,\ref{fig:phasediag_exp}(c)]. 
We notice, on passing, that the phase diagram of hole-doped La-based 
cuprates is characterized by a considerable (and unavoidable) coexistence of different 
structural phases, i.e., the low-temperature tetragonal (LTT) 
and the low-temperature orthorombic (LTO) phases\cite{tidey2022pronounced}.
Here, the assigned filamentary phase is more prominent, which can be understood 
as the effect of higher disorder (cf. with Fig.\,4(a) in Ref.~\cite{Leridon2020}). 

Notice that in panels (a) and (b) of Fig.\,\ref{fig:phasediag_exp}, $T_c$ evolves rapidly with the control parameter, while the CO temperature is approximately constant. The situation reverses in panel 
(c). Clearly, this is due to the different way the control parameter couples to the two competing phases. In (a) and (b) the magnetic field destabilizes the superconducting phase having little influence on the CO. Instead, in (c) the structural distortion stabilizes the CO phase and has little effect on the superconducting $T_c$. In the case of $^4$He, pressure stabilizes the solid phase, which indeed has a larger critical temperature slope.

Summarizing, the cuprate phase diagrams are in many respects similar to the 
$^4$He  phase diagram shown in Fig.\,\ref{fig:phasediag_exp}(d), namely: i) similar 
critical temperatures for SC/superfluid and CO/solid, ii) vertical transition lines and iii) 
FSC/supersolid phase induced by disorder and grain boundary effects. 

For $^4$He, the control parameter is the pressure $P$. At low temperatures, 
and at a critical pressure nearly independent of the temperature, the superfluid 
phase transforms into the solid phase. The analogy between a continuum 
system like $^4$He and lattice systems as cuprates is not new. 
Indeed, at least from a theoretical point of view, there is a long 
tradition~\cite{zilsel1965pseudospin, liu1973hard} of modelling $^4$He on discrete lattices, very similar in spirit to the model we shall be using for cuprates 
in this work. Also, analogies between the phase diagram of cuprates and $^4$He had been emphasized before \cite{Uemura2004}.

The vertical transition line 
$T_{\text{SF}\leftrightarrow \text{S}}(P)$ in (c) between the superfluid (subscript SF) and 
the solid phase (subscript S), more often drawn as horizontal with exchanged axes, was 
one of the first arguments put forward by 
London to advocate for some form of condensation in the early days 
of research on superfluids. Indeed, according to the Clausius-Clapeyron 
equation, relating the changes in entropy $\Delta S$ and in volume 
$\Delta V$ at a first-order phase transition, the divergent derivative 
$\mathrm dT_{\text{SF}\leftrightarrow \text{S}}/\mathrm dP=\Delta V/ \Delta S $ 
implies that $\Delta S=0$, thus identifying the superfluid as a 
practically zero-entropy state, like{ a solid phase} \cite{London1954}.
Analogously, the divergent 
$\mathrm dT_{\text{SC}\leftrightarrow \text{CO}}/\mathrm dH=\Delta M/ \Delta S $ 
(with $\Delta M$, the change in magnetization) 
in cuprates implies that at the critical field, the same quasiparticles flip 
from a momentum-condensed state to a real-space condensed state, that represents 
two equally ordered (low-entropy) states.  

A minimal model to investigate the instability that can occur in the 
particle-hole or particle-particle channel, i.e., pre-formed electron 
pairs that are paired in real space (precursors of CO) 
or in momentum space (Cooper pairs, precursors of SC) 
is the two-dimensional attractive Hubbard model \cite{Micnas1990}.
This model enjoys the property that, exactly at half-filling (one 
electron per unit cell), CO and SC are 
degenerate. Moving the electron density away from half-filling usually 
tilts the balance in favour of SC, unless other 
interactions (e.g., a nearest-neighbor repulsion) are present. 

The attractive Hubbard model can be mapped onto the repulsive Hubbard 
model \cite{Micnas1990}, which has a spin-density-wave ground state. In 
this representation, spin-density-wave order along $z$ describes a 
charge-ordered state, while order in the $xy$ plane describes
SC. Order along any other direction 
maps into uniform ``supersolid"  order,  where SC and CO coexist. 
Since the free energy of the repulsive model is invariant with respect to $O(3)$ rotations of the order parameter, one concludes that the
charge-ordered, superconducting  and supersolid phases are degenerate.
In two spatial dimensions, this suppresses 
the ordering temperature to $T=0$, due to the Mermin-Wagner 
theorem \cite{mermin1966absence}.

More insight into the competition between CO and 
SC can be gained considering the limit of a large Hubbard coupling, where the model can be mapped onto a Heisenberg 
model~\cite{Micnas1990} of interacting pseudospins (analogous to Anderson's 
pseudospins). Here, the pseudospin projection along $z$, up or down, 
encodes a double occupied or an empty site on the lattice, respectively, while 
the in-plane component encodes superconducting correlations.  
While the Hubbard model is genuinely quantum, it is well
known\cite{Chakravarty1989} that, provided the ground state is ordered 
above a microscopic scale (the Josephson correlation length) 
$\xi_\text{J}\approx \hbar c/J_\text{s}$, the system can be described by a 
classical theory. Here, $J_\text{s}$ 
and $c$ are the stiffness and 
the zero-temperature spin-wave velocity of the effective Heisenberg model, 
respectively. Using estimates appropriate for the 
latter \cite{Chakravarty1989}, one obtains 
$\xi_\text{J}=2\pi a/C_\text{s}\approx 10.9\, a$ with $C_\text{s}$ 
a constant, and $a$ 
the lattice spacing. Therefore, we can use a classical effective lattice spin model to study the competition between CO and
SC. Each pseudospin in the lattice model represents a 
cluster of elementary unit cells
with a linear dimension of order $\xi_\text{J}$ behaving
as a classical variable. The superconducting transition in two dimensions belongs to the 
Berezinskii--Kosterlitz--Thouless (BKT) universality class, 
and is characterized by the appearance of a finite stiffness and the binding of 
vortices and antivortices.

While the above scenario is very appealing to formulating a statistical mechanical description of the competition between CO and SC, a full $O(3)$ symmetry is clearly a drawback of the 
model. Indeed, for the repulsive Hubbard model, this is a consequence of rotational invariance, but there is no such fundamental symmetry in a generic attractive model. One expects that CO 
and SC can be tuned to an approximate $O(3)$ symmetry point by a non-ordering field  (e.g.$p$ for cuprates), but there is 
no reason why the barrier between these states should vanish at  
the $O(3)$ point.
In other words, the $O(3)$ symmetry is only approximate, in the sense that the charge-ordered
and superconducting phases are still degenerate but are separated by barriers. 

Based on these considerations, we study an effective classical spin model on a square lattice, with nearest neighbour exchange interaction and three relevant parameters: an exchange anisotropy, to tilt the balance between easy-axis (charge) and easy-plane (superconducting) order; a potential barrier, to remove the unphysical high degeneracy of the $O(3)$ symmetric point; a random field to mimic disorder. 

In the clean system (without disorder), 
we find that the presence of the barrier allows for a finite-temperature phase transition, otherwise forbidden at the $O(3)$ point. Once disorder is taken into account, CO is fragmented into different domains, 
resulting in a polycrystalline charge-ordered phase, and FSC sets in as a parasitic phase at the 
domain boundaries \cite{caprara2020doping,Leridon2020}.

Our analysis is carried out by means of Monte Carlo (MC) calculations, which 
allow us to study not only the ground state, as in Ref.\,\cite{Leridon2020}, 
but the thermodynamic phase diagram itself and the behaviour in temperature of 
the various physical quantities.

The scheme of this work is the following. In 
Sec.\,\ref{sec:theory}, we discuss the model and methods of 
investigation. In Sec.\,\ref{sec:barrier}, we discuss the 
properties of the model in the absence of disorder, highlighting
the role of the potential barrier and using the result of the model 
in the absence of a 
barrier\cite{lee2005helicity, cuccoli1994critical, cuccoli1995two} 
as a benchmark. The phase diagrams for small and large 
values of the potential barrier are discussed in detail in 
Secs.\,\ref{b01} and \ref{b1}, respectively.
In Sec.\,\ref{dirty}, we include the effect of disorder and show that 
this is crucial to promote FSC. Our 
concluding remarks are found in Sec.\,\ref{concl}.


\section{Model and Methods}
\label{sec:theory}

Above the Josephson scale, we can model our system with a classical order 
parameter. Therefore, 
as in Ref.~\cite{attanasi2009competition, Leridon2020, caprara2020doping} 
we consider a coarse-grained model of classical pseudospin vectors 
$\vecs_{\bm R}$ on the sites ${\bm R}$ on a square lattice, each representing a 
region of area $\xi_\text{J}^2$ of the quantum system. The new (coarse-grained)
lattice spacing is set as $a'=1$ and the linear size is $L$ (i.e., the lattice hosts 
$N=L^2$ sites), with periodic boundary conditions. The states with 
positive or negative pseudomagnetization along the $z$ axis represents two 
variants of the charge-ordered state, related in the original quantum microscopic model by a translation symmetry, while the in-plane 
pseudomagnetization describes the superconducting state 
\cite{attanasi2009competition, Leridon2020, caprara2020doping}. 
In order to lighten the notation, we will henceforth refer to the 
pseudomagnetization simply as magnetization, not to be confused with the physical magnetization mentioned in connection with the Clausius-Clapeyron argument above.  
In the following, we set $|\vecs_{\bm R}|=1$, and fix the reference 
frame so that the three Cartesian components of the vector 
$\vecs_{\bm R}$ are $S_{\bm R}^x=\sin\varphi_{\bm R}\cos\theta_{\bm R}$, 
$S_{\bm R}^y=\sin\varphi_{\bm R}\sin\theta_{\bm R}$, 
and $S_{\bm R}^z=\cos\varphi_{\bm R}$, in terms of the polar and
azimuthal angles $\theta_{\bm R}$ and $\varphi_{\bm R}$.
$\theta_{\bm R}$ can be identified with the phase of the superconducting 
order parameter. 

The competition between CO and SC is captured by 
the classical anisotropic Heisenberg model (XXZ model) with an effective 
barrier potential term and a random field mimicking disorder,
\begin{equation}
H=- J\sum_{\langle\bm R,\bm R'\rangle} \left(S^{x}_{\bm R} 
S^{x}_{\bm R'} +S^{y}_{\bm R} S^{y}_{\bm R'}  
+\a\, S_{\bm R}^z S_{\bm R'}^z \right) +
4B \sum_{\bm R} \left(S^z_{\bm R}\right)^2 \left[1
-\left(S^z_{\bm R}\right)^2\right]  
+\frac{W}{2}\sum_{\bm R} h_{\bm R} S^z_{\bm R},
\label{eq:XXZ_Htot}
\end{equation}
where the symbol $\langle{\bm R},{\bm R'}\rangle$ specifies that the 
sum runs over nearest-neighbouring sites. We fix the interaction 
strength $J=1$ unless otherwise specified. Furthermore, we use the anisotropy parameter $\a\ge 0$ to tune the 
ground state from being superconducting to charge-ordered. This corresponds to keep constant the ground-state energy of the SC and tune the one of the CO i.e. a CO-driven transition. As we shall see, a simple rescaling of the energy units allows describing a SC-driven transition.

The second term in Eq~\ref{eq:XXZ_Htot}
is the barrier potential, whose height is adjusted by the parameter 
$B$. Its role, as  anticipated in Sec.\,\ref{intro}, is to eliminate 
the unphysical degeneracy of the charge-ordered and superconducting 
state with all possible intermediate supersolid phases for $\alpha=1$.
In the following, we will still call the 
$\alpha=1$ case ``the $O(3)$" or ``isotropic" point, keeping in mind that such 
terminology refers only 
to the first term of the Hamiltonian in Eq.~\ref{eq:XXZ_Htot}.

The last term 
in Eq.\,(\ref{eq:XXZ_Htot}) is a random field that mimics impurities 
coupled to the charge density in 
a real system. We take $h_{\bm R}$ as independent random variables with a 
flat probability distribution between $-1$ and $+1$. The strength of disorder is controlled by the parameter $W$. As we shall show, this term is crucial to promote the polycrystalline behaviour of CO for $\a\gg 1$, as well as the occurrence of FSC in a certain range of anisotropy $\a\gtrsim 1$, in the form of topologically protected domain walls between regions hosting two different realizations of CO. 

In the case where $B=W=0$, Eq.\,\eqref{eq:XXZ_Htot} is the bare XXZ model which
has been widely studied {in the literature}~\cite{lee2005helicity, cuccoli1994critical, cuccoli1995two},
and whose phase diagram will be used as a benchmark case when discussing the effect of the energy barrier. In the bare model,
the anisotropy $\a$ allows switching from the BKT universality class, {for $\a<1$, where the ground state is superconducting,} 
to the Ising universality class, {for $\a>1$, where it is a charge-ordered ground state.} 
{Finally,} in the isotropic limit $\a\rightarrow 1^{+,-}$, the critical temperature goes to zero logarithmically \cite{menezes1992calculation}, and at $\a=1$ no finite-temperature phase transition is possible, according to the Mermin-Wagner theorem \cite{mermin1966absence}.

{The  presence of a finite energy barrier separating the three equivalent ground states $\varphi_{\bm R}=0,\p/2,\p$ (i.e., $S_{\bm R}^z=-1,0,1$) makes the model} 
no longer invariant with respect to $O(3)$ rotations of the order parameter, and the Mermin-Wagner theorem does not apply. Indeed, we find that{, for $B > 0$,} the ordering temperature remains finite for all $\a$. Furthermore, since the effect of the barrier persists at finite temperatures, metastability regions appear in the resulting $T$\,vs\,$\alpha$ phase diagram. 

In order to study the physical quantities related to the effective Hamiltonian, Eq.\,\eqref{eq:XXZ_Htot}, as functions of the temperature, we performed large-scale Monte Carlo (MC) simulations{, with systems of linear size $L$} ranging from 
$L=16$ up to $L=256$. We used the Metropolis and simulated annealing algorithms to optimize the thermalization process: at the highest temperature reached in our calculations the system evolves from an initial configuration of random pseudospins until it reaches its equilibrium state, then the temperature is slightly decreased and a 
new thermalization starts from the final configuration of the previous step. This process is iterated until the lowest temperature of interest 
is reached. At each Metropolis step, the whole lattice is updated according to the Metropolis prescription \cite{metropolis1953equation},
either sequentially updating all the $L\times L$ pseudospins or by $L\times L$ 
random choices of the pseudospin to be updated. Thermal averages of any 
observable $\mathcal O$ are obtained as the average over $N_\text{MC}$ 
(at least $10^3$) measures, 
\[
\langle \mathcal{O} \rangle =\frac{1}{N_\text{MC}} 
\sum^{N_{\text{MC}}}_{j=1} \mathcal{O}_j, 
\]
taken $\tau_\text{MC}$ Metropolis steps apart from one another, 
$\tau_\text{MC}$ being of the order or larger than the autocorrelation time 
(for the clean system, 
typically we take $\tau_\text{MC}=30- 100$ Metropolis steps). To account for the thermalization time, we finally discard the initial 
$N_{\text{out}}$ (at least $10^5$) Metropolis steps. 
In the presence of the random field, we also average over 
$N_{\text{dis}}$ realizations of the disorder (henceforth, this average
is marked by an overline).

\subsection{Physical observables}
\label{sec:physobs}

To analyse the tendency to order, we compute the mean-square magnetization, 
\begin{equation}
\widetilde\chi^\nu\equiv N \< m_\nu^2 \> =\frac{1}{N}\left\langle 
\left(\sum_{\bm R}S_{\bm R}^\nu\right)^2 \right\rangle,
\quad \nu=x,y,z,
\label{eq:XXZ_chinn}
\end{equation} 
where $m_\nu =\frac{1}{N} \sum_{\bm R} S_{\bm R}^\nu $ is the 
magnetization per unit surface area calculated at each MC step. Note 
that the mean-square magnetization is directly related to the charge ($\nu=z$) and 
superconducting $\nu=x,y$ 
susceptibilities\cite{cuccoli1995two},
\[
    \chi^\nu = \frac{1}{T}\left[ \left\< 
    \left(\sum_{\bm R} S_{\bm R}^\nu\right)^2\right\> 
    - \left\langle \sum_{\bm R} S_{\bm R}^\nu \right\rangle^2 
    \right].
\]
In the absence of long-range order
$\<m_\nu\>=0$, therefore, for a BKT system in the thermodynamic limit, 
$\chi^\nu=N\widetilde\chi^\nu/T$. {Of course, in numerical calculations, 
the system is always finite and never reaches the real thermodynamic limit, 
preventing the vanishing of $\<m_{x,y}\>$.} In the following, we will use the quantity
$\widetilde\chi^{xy}\equiv \frac{1}{2} (\widetilde\chi^x +\widetilde\chi^y)$, 
to monitor the superconducting correlations. While $\widetilde\chi^{xy}$
 can be seen as a proxy of the superconducting susceptibility, 
in the charge-ordered
phase, instead, the order parameter is nonzero, so that $\chi^z$ and 
$\widetilde\chi^z$ are not simply proportional. In this case, 
to monitor the response of the CO correlations, we use both 
the susceptibility $\chi^z$, which is the true response of the system to an external field, 
and $\widetilde\chi^z$.

In order to assess a global BKT superconducting transition, 
we compute the in-plane superfluid stiffness $J_\text{s}$, associated with 
the superconducting phase rigidity and  defined as the second derivative of the free energy with respect to a twist of the SC phase angle $\delta \theta$, e.g., along the $x$ direction:
\[
\begin{split}
    J_\text{s}(L, T) & =-T \frac{\partial^2 \ln Z(\delta\theta)}
    {\partial\delta\theta^2}\Big|_{\delta\theta=0}=\\
    & =    \frac{1}{L^2}\left\langle \sum_{{\bm R}}  
    \sin\varphi_{\bm R} \sin\varphi_{{\bm R}+\hat{\bm x}} 
\cos(\theta_{\bm R} - \theta_{{\bm R}+\hat{\bm x}}) \right\rangle \\
    & \, -\frac{1}{L^2 T}
\left\langle \left(\sum_{\bm R} \sin\varphi_{\bm R} 
\sin\varphi_{{\bm R}+\hat{\bm x}} \sin(\theta_{\bm R} 
- \theta_{{\bm R}+\hat{\bm x}})\right)^2 \right\rangle 
\end{split}
\]
where $Z$ is the partition function and $\hat{\bm x}$ is the unit vector 
in the $x$ direction.

To perform the extrapolation to the thermodynamic limit of the BKT 
critical point we use the BKT scaling of the superfluid 
stiffness \cite{weber1988monte},  
\begin{equation}
\frac{J_\text{s}(L,T_\text{BKT})}{1+\left[2\ln(L/L_0)\right]^{-1}}=\frac{2}{\p}T_\text{BKT},
\label{eq:XXZ_scalingBKT}
\end{equation}
where $T_\text{BKT}$ is the BKT critical temperature  and we take $L_0$ as a 
fitting parameter. 

Whenever needed, we complement our analysis of the superconducting 
state, with a closer inspection into the occurrence of vortices in
the pattern of the local superconducting order parameter 
(the in-plane magnetization).
A vortex (anti-vortex) is identified whenever a variation of 
$2\pi$ ($-2\pi$) of the superconducting phase $\theta_{\bm R}$ is found in a 
closed path around a single plaquette with side equal to
the lattice spacing. 
Defining the superconducting phase difference at site $\bm R$ in the direction 
of the $\hat{\boldsymbol\nu}$ unit vector 
($\hat{\boldsymbol\nu}=\hat{\bm x},\hat{\bm y}$) as 
\[
\Theta_{\hat{\boldsymbol\nu}}(\bm R)=[ \theta_{\bm R} - 
\theta_{\bm R+\hat{\boldsymbol\nu}} ]_{-\pi}^{+\pi},
\]
the notation $[\cdot]_{-\pi}^{+\pi}$ meaning that we take the value 
modulus $2\pi$ so that $\Theta_{\hat{\boldsymbol\nu}}(\bm R) 
\in (-\pi, \pi ]$,
the circulation of the superconducting phase around a plaquette whose center 
is located at $\bm R +\frac{1}{2}(\hat{\bm x}+\hat{\bm y})$ is
\[
\Theta_{\hat{\bm x}}(\bm R) + \Theta_{\hat{\bm y}}(\bm R+ \hat{\bm x}) - 
\Theta_{\hat{\bm x}}(\bm R+\hat{\bm y}) - \Theta_{\hat{\bm y}}(\bm R) = 
2\pi n_{\bm R},
\]
where $n_{\bm R}=\pm1$ is the integer vorticity in the phase angle 
$\theta_{\bm R}$ going around the plaquette \cite{teitel2013two}.
Summing over all positive (negative) vorticities per unit length we 
obtain the density of vortices (antivortices) 
$\rho_\text{V}>0$ ($\rho_\text{AV}<0$), defining the total vorticity as 
\begin{equation}
    \rho_\text{V, tot}=\rho_\text{V}-\rho_\text{AV}.
    \label{eq:vorticity}
\end{equation}

Concerning the charge-ordered state, we define $T_\text{CO}$ using the crossing 
point (as a function of temperature) of the kurtosis of the pseudospin distribution function, i.e., the Binder 
cumulant 
\begin{equation}
    U_N=1-\frac{\langle m_z^4 \rangle}{3\langle m_z^2\rangle^2}.
    \label{eq:UN}
\end{equation} 
Its value at the critical temperature is indeed less sensitive to finite-size effects, as compared to the CO order 
parameter $\langle m_z\rangle$, and unbiased by fitting functions 
and \emph{a priori} scaling hypotheses. In the thermodynamic limit $N\ra\infty$, 
one expects $U_N\ra 0$ in the high-temperature limit, while $U_N\ra 2/3$ in 
the ordered phase for $T\ra 0$ \cite{binder1981finite, landau1983non, 
hasenbusch2008binder}.

\section{Clean system}
\label{sec:barrier}

\subsection{Metastability and spinodal lines}
\label{sec:meta}

{Let us start discussing the MC numerical results starting from the clean case in the presence of a finite energy barrier, i.e. $B>0$.}
As can be expected, one prominent effect of the barrier is to introduce metastability 
in the system. Already at zero temperature, there exists a range of values of 
$\a^*_\text{CO}(B)<\alpha< \a^*_\text{SC}(B)$ where both phases are local minima 
of the energy \cite{venditti2023charge}. 
The spinodal points $\a^*_\text{CO,SC}(B)$ mark the limit of stability of the less 
stable phase (i.e., CO, decreasing $\alpha$). 

At zero temperature, the spinodal points $\a^*_\text{CO,SC}(B)$ can be calculated 
analytically as a function of the barrier height $B$. Indeed, increasing $\a$ at 
$T=0$, we can assume that all pseudospins are parallel, i.e., $\varphi_{\bm R}=\varphi$ 
and $\theta_{\bm R}-\theta_{\bm R'}=0$. Up to constant terms in the angle $\varphi$, 
the total energy per site
from Eq.\,\eqref{eq:XXZ_Htot} takes the simplified form
\begin{equation}
\frac{1}{N}E(\varphi)=(1-\alpha)\cos(2\varphi) - \frac{1}{2} B \cos(4\varphi).
\label{eq:H(phi)_T=0}
\end{equation}

\begin{figure}[t]
    \centering
    \includegraphics[width=0.9\linewidth]{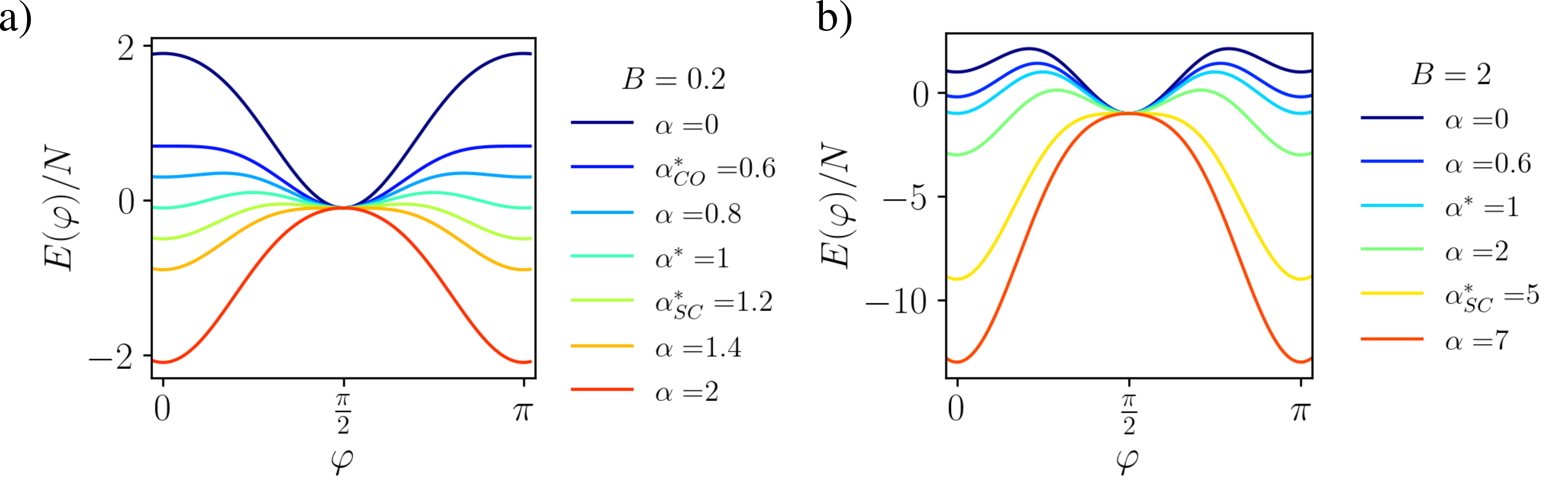}
    \caption{Energy landscapes [Eq.\,\eqref{eq:H(phi)_T=0}] at $T=0$, for  various $\a$ and: 
    a) $B=0.2$; b) $B=2$. Notice that in this case even for $\alpha=0$ the charge-ordered phase remains 
    as a metastable minimum, and there are two possible realizations (two equivalent 
    minima).}
    \label{fig:H_T=0}
\end{figure}

The resulting energy landscapes for $B=0.2$ and $B=2$, with varying $\alpha$, are given in 
Figs.\,\ref{fig:H_T=0}(a) and (b), respectively.
The free energy has a single 
superconducting global minimum at $\varphi=\pi/2$, up to some $\a=\a_\text{CO}^*(B)$, 
after which two new \emph{local} (metastable) minima at $\varphi=0,\pi$, corresponding to 
CO, appear. Crossing the isotropic point $\a^*$ the situation gets reversed: the new global minima are found at $\varphi=0,\pi$ and the $\varphi=\pi/2$ 
configuration becomes a local (metastable) minimum, disappearing at some $\a^*_\text{SC}(B)$, 
after which only the two equivalent charge-ordered states survive.

By substituting $\varphi=\pi/2$ and $\varphi=0,\p$ in the second derivative 
of Eq.\,\eqref{eq:H(phi)_T=0}, we obtain the two spinodal points at zero 
temperature, 
\begin{equation}
\label{eq:a*}
\a^*_\text{SC,CO}(B)=1\pm 2B.
\end{equation}
For $B=0$ the model becomes fully symmetric and the zero-temperature spinodal points merge 
into the zero-temperature transition point. 

The study of metastable states at finite temperatures is particularly 
challenging, both in real experiments and in numerical simulations. If  
the system is prepared in a metastable state, there is a finite probability 
that a bubble of the more stable phase nucleates in a finite time and then grows. 
Therefore, sooner or later the system will transit to the more stable phase.  
As a consequence, spinodal lines, marking the limit of stability of the metastable phase, require time-domain considerations to be defined.
A metastable phase is well-defined only if it persists at least for its equilibration time, otherwise, it can not be considered a thermodynamic phase. 
The line where the equilibration and the nucleation times become equal defines the dynamical spinodal line~\cite{Cavagna2009}. While these considerations allow a rigorous definition of spinodals, here we circumvent the problem of estimating these times and take a more pragmatic approach which is enough for our purpose, as follows. 
For a small enough system, during a simulation, the system may get trapped in one 
metastable state for a long time (compared to the equilibration time of the state) 
and sporadically change to a different state, where it gets trapped again. Evolving 
the system long enough, we can construct a reliable effective 
probability density function $P_{\text{eff}}(m_z)$ for the order 
parameter associated with CO, that is encoded in $m_z=\frac{1}{N}\sum_{\bm R} S^z_{\bm R}$. By definition, 
the free energy of the system is
$F(m_z)=-T\ln[P_{\text{eff}}(m_z)/\sum_{m_z} P_{\text{eff}}(m_z)]$.
Notice that by doing the histogram in $m_z$ one automatically takes into 
account the correct measure for the probability distribution. It is 
easy to check that, for $B=0$, the probability distribution is flat at $\alpha=1$, 
$P(m)=\frac{1}{2}$, consistent with a flat free energy.   

The numerical identification of the spinodal lines and the equilibrium coexistence line of SC and CO can be obtained by studying the form of $F(m_z)$ as a function of the parameters. For a fixed 
value of $\a$, the first-order transition temperature $T_{SC\leftrightarrow CO}$, 
is defined as the temperature at which the absolute minimum of the free energy 
changes from $m_z=\pm 1$ to $m_z=0$. A spinodal temperature $T_\text{sp}$ is defined 
by the local minimum of a metastable phase becoming an inflection point. Thus, 
once we extracted all the effective free energies $F(m_z)$ at each temperature, we can infer the full phase diagram.  

Since the flip of the whole phase is a very rare event, we need to take a very 
small system in order to have enough flips to consider the system at equilibrium 
within a reasonable simulation time. As a proof of principle and in order to have an approximate map of 
the spinodal lines, we take $L=4$ and construct $P_{\text{eff}}(m_z)$ using 
histograms of $m_z$ measured at each MC step. 
The system is evolved for
$N_\text{MC}=5\times 10^5$ after $N_\text{out}=5\times 10^5$, with $\tau_\text{MC}=50$. Of course, it must be borne in mind that the condition $m_z=0$ 
cannot distinguish a superconducting state from a charge-disordered 
state. To construct the phase diagram one has to complement the previous study with
the computation of the superfluid stiffness.

The search for metastable states and first-order lines  within the 
above protocol becomes harder and harder with increasing $B$. Thus, the investigation of the 
metastable states must be adapted to the cases of a small ($B\ll 1$)
and a large ($B\gg 1$) barrier. Henceforth, we shall refer to the paradigmatic cases
$B=0.2$ and $B=2$ to discuss the two different regimes. 

In the case of large $B$, we follow a different protocol to numerically estimate 
the spinodal points. For $\a<\a_\text{B}$ ($\a>\a_\text{B}$) the superfluid 
stiffness (Binder cumulant) are computed starting from a charge-ordered 
(superconducting) configuration and heating up 
the system, using the simulated annealing algorithm.
We define a spinodal point using the temperature at which 
$J_\text{s}$ or $\<m_z^2\>$ jump to their finite value, checking that 
this temperature is not strongly dependent on the system size $L$. 
The absence of a significant size dependence can also be viewed as a confirmation of the spinodal points extracted with the free energy protocol.

In Sec.\,\ref{bic} we characterize bicriticality within our model, whereas in Secs.\,\ref{b01} and \ref{b1} we discuss in detail two paradigmatic examples of phase diagrams, for a small and large value of the barrier height, respectively.

\subsection{Bicriticality}
\label{bic}

According to the Mermin-Wagner theorem~\cite{mermin1966absence}, the isotropic 
Heisenberg model, i.e., Hamiltonian\,\eqref{eq:XXZ_Htot} with $W=B=0$ and 
$\alpha=1$ has no long-range order. We discussed already in Sec.\,\ref{sec:theory} the 
unphysical nature of the full $O(3)$ symmetry that 
characterizes the isotropic point without a barrier. 
We will show in Secs.\,\ref{b01} and \ref{b1} that the bicritical point $(\a_\text{B},T_\text{B})$,  
at which SC, CO and the disordered phase meet, gets shifted to $\a_\text{B}>1$. 

\begin{figure}
	\centering
    \includegraphics[width=0.8\linewidth]{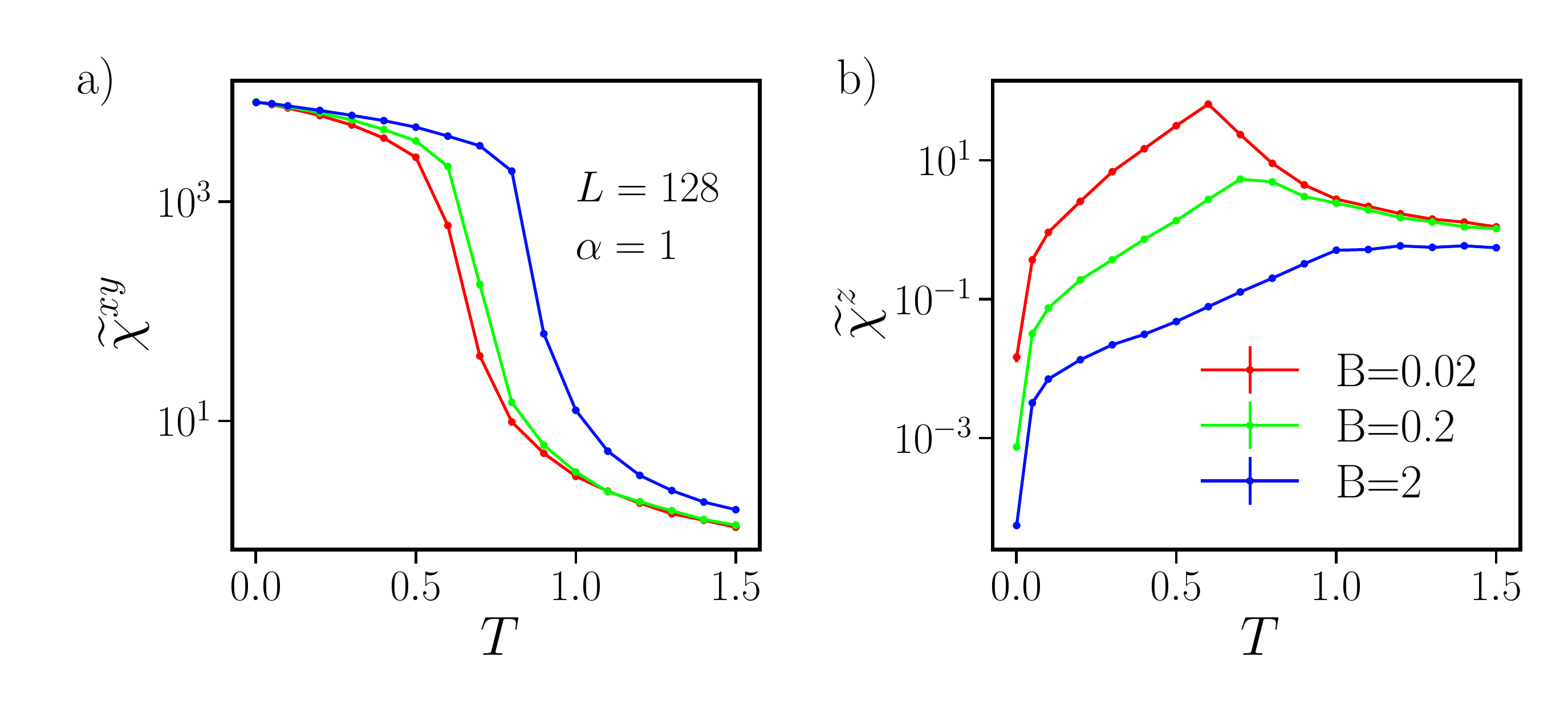}
    \caption{a) Superconducting $\widetilde\chi^{xy}$ and b) charge-ordered $\widetilde\chi^z$ mean-square 
    magnetizations vs. temperature $T$ for the isotropic case $\a=1$ 
 at various heights of the potential barrier $B$. While for the bare Heisenberg model 
 ($B=0$) no transition is possible, the presence of a barrier allows for a BKT 
 transition at $\a=1$. We used 
$L=128$, $N_{MC}=2\times 10^3 $, $N_{out}=5\times 10^4$ and  $\tau_\text{MC}=100$.  
The error bars are calculated using the bootstrap resampling method with 100 datasets and 
blocks of size 100 \cite{efron1992bootstrap}.}
	\label{fig:chi_a1}
\end{figure}

To gain a first insight into this phenomenon, we  
compare in Fig.\,\ref{fig:chi_a1}
the {superconducting} $\widetilde\chi^{xy}$ and {charge-ordered} $\widetilde\chi^{z}$ functions, 
defined as in Eq.\,\eqref{eq:XXZ_chinn} for different values of $B> 0$. 
When the barrier is present ($B>0$), we observe a sizable
$\widetilde\chi^{xy}$. The temperature at which the superconducting response significantly 
rises is an increasing function of $B$ (see 
Fig.\,\ref{fig:chi_a1}a). At lower temperatures, the mean-square magnetization tends to a finite value, 
indicating the stabilization of superconducting correlations. Indeed, contextually, $\widetilde\chi^z$ presents a peak and is driven to zero at low temperatures. This behaviour is characteristic of the bare XXZ model with $\alpha< 1$, i.e., in the superfluid region of the phase diagram. In the presence of the 
barrier, we find that the same results also persist for a small range of $\alpha> 1$.
Thus, the effect of the barrier is to shift the bicritical point ($\a_\text{B},T_\text{B}$)
to $\alpha_\text{B}> 1$. However, as discussed in the previous section, at $T=0$ the superconducting and the 
charge-ordered phases are degenerate at $\alpha= 1$. This implies that the first order line $T_{CO\leftrightarrow SC}$, which by definition starts at   
{$(\alpha=\a^*\equiv 1$, $T=0)$} and ends at the bicrtical point, must have a positive slope. This indicates 
that for a small range $\alpha\gtrsim 1$ and lowering the temperature, one has the sequence of phases: disorder 
$\rightarrow$ SC $\rightarrow$ CO. 
Thereby, two spinodal lines starting from $(\a_\text{B},T_\text{B})$ 
and terminating at points $\a^*_\text{CO,SC}(B)$ 
and $T=0$ should appear, as will be illustrated in the next sections. 

\subsection{Phase diagram for $B=0.2$}
\label{b01}

 \begin{figure}[t]
    \centering
    \large{$B=0.2$\\}   
    \includegraphics[width=\linewidth]{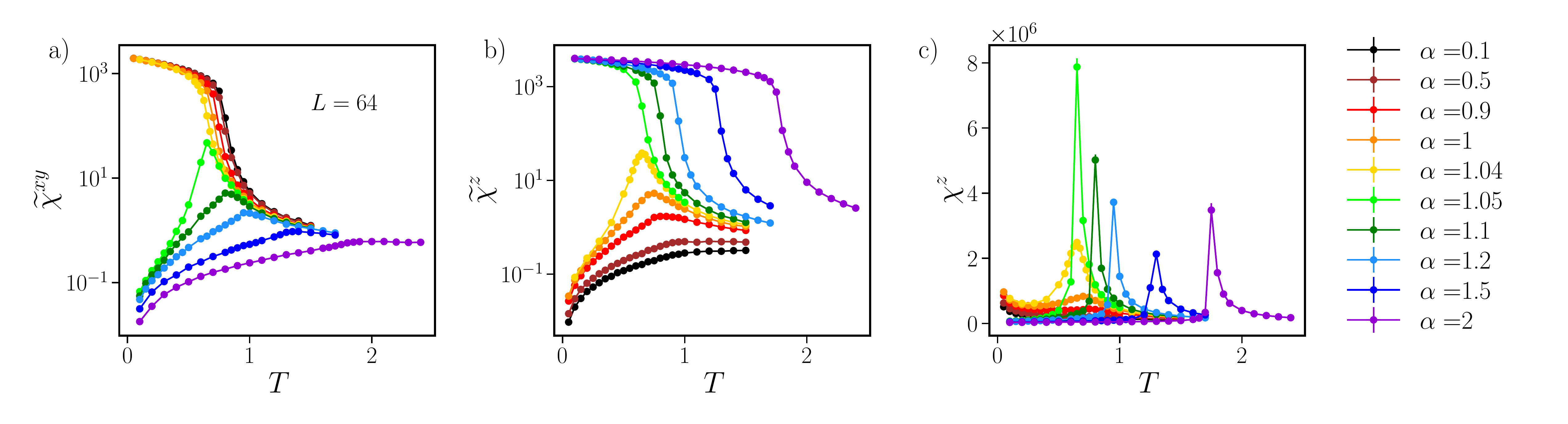}\\
    \large{$B=2$\\}
    \includegraphics[width=\linewidth]{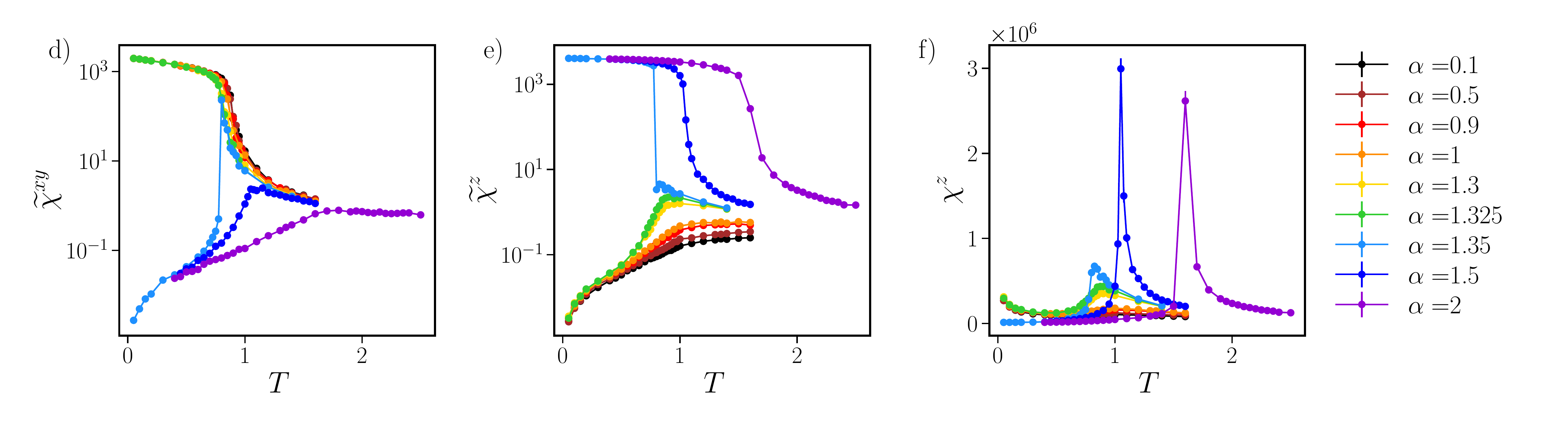}
    \caption{
    a) $\widetilde\chi^{xy}$, b) $\widetilde\chi^{z}$, and c) $\chi^z$, at different values of the anisotropy parameter 
    $\alpha$, for a barrier parameter $B=0.2$.
    We used $L=64$, $N_{MC}=10^4$ and $\tau_\text{MC}=30$ (50) discarding the first 
    $6\times 10^5$ ($2.5\times 10^6$) MC configurations when $\alpha\leq 1$ ($\a>1$).    
    d) $\widetilde\chi^{xy}$, e) $\widetilde\chi^{z}$, and f) $\chi^z$ at different values of the anisotropy parameter 
    $\alpha$, for a barrier parameter $B=2$. Parameters are the same except that $\tau_\text{MC}=40$ and 
    $N_\mathrm{out}=8\times 10^5$.
    The error bars are calculated using the bootstrap resampling method with 100 
    dataset and blocks of size 100.
    }
    \label{fig:Nmsq_b0.1}
\end{figure}

In this section, we discuss the case $B=0.2$. 
In Fig.\,\ref{fig:Nmsq_b0.1} we plot the {superconducting} $\widetilde\chi^{xy}$ (panel a)
and {charge-ordered} $\widetilde\chi^z$ (panel b) mean-square magnetization, as well as the susceptibility
$\chi^z$ (panel c), for different values of the anisotropy parameter, in the range 
$0.1<\a<2$. 

The {superconducting} mean-square magnetization $\widetilde\chi^{xy}$ 
grows monotonously by 
lowering the temperature for values of the anisotropy parameter as large as $\a=1.04$, i.e. above the isotropic Heisenberg limit. 
For the same range of anisotropy, 
$\widetilde\chi^{z}$ shows a maximum and then drops nearly to zero at lower temperatures. Clearly, thus, 
in this region the superconducting phase prevails at low temperature, {so that, at some temperature $T_{\text{BKT}}$, the system transitions from a high-temperature disordered state 
to a superconducting state.}

For $\a>1.04$, the situation gets reversed with the charge correlations growing monotonically and the 
superconducting ones getting suppressed. 
This behaviour is coherent with the results found in the anisotropic Heisenberg model 
without a barrier ($B=0$), where $\widetilde\chi^{z}$ decreases with $T$ for $\a\ll 1$ while for 
$\alpha\lesssim 1$ it displays a peak at $T\simeq T_\text{BKT}$\cite{cuccoli1995two},
as a precursor of the Ising transition that is found for $\a>1$. 

Observing the yellow curve corresponding to $\a=1.04$, in Fig.\ref{fig:Nmsq_b0.1}a, Fig.\ref{fig:Nmsq_b0.1}b, and Fig.\ref{fig:Nmsq_b0.1}c, one
can understand the importance of considering all three quantities $\widetilde\chi^{xy},\,\widetilde\chi^z$ 
and $\chi^z$. As a matter of fact, the CO susceptibility $\chi^z$ in Fig.\ref{fig:Nmsq_b0.1}c presents a peak at $\a=1.04$, although 
smeared with respect to the peaks for $\a\geq1.05$; concurrently, $\widetilde\chi^{xy}$, Fig.\ref{fig:Nmsq_b0.1}a, shows that a 
superconducting state is present at $\a=1.04$. The doubt about whether the system has a superconducting or charge-ordered ground state is solved by looking at $\widetilde\chi^z$, Fig.\ref{fig:Nmsq_b0.1}b, in which the $\a=1.04$ curve grows with $T$ following 
the typical behaviours of the charge-ordered states $\a\geq 1.05$, but then decreases below a temperature 
$\widetilde T=0.65$, at which $\widetilde\chi^z\sim 37$. 
 The BKT scaling of $J_s$ for $\a=1.04$ allowed us to extract $T_{\text{BKT}}=0.575\lesssim\widetilde T$.

As soon as $\a\geq 1.05$, the main response of the system is in the out-of-plane direction (corresponding to 
CO), as it is clear looking 
at $\widetilde\chi^z$, Fig.\ref{fig:Nmsq_b0.1}b, and at the susceptibility $\chi^z$, Fig.\ref{fig:Nmsq_b0.1}c.

\begin{figure}
    \centering
    \includegraphics[width=0.85\linewidth]{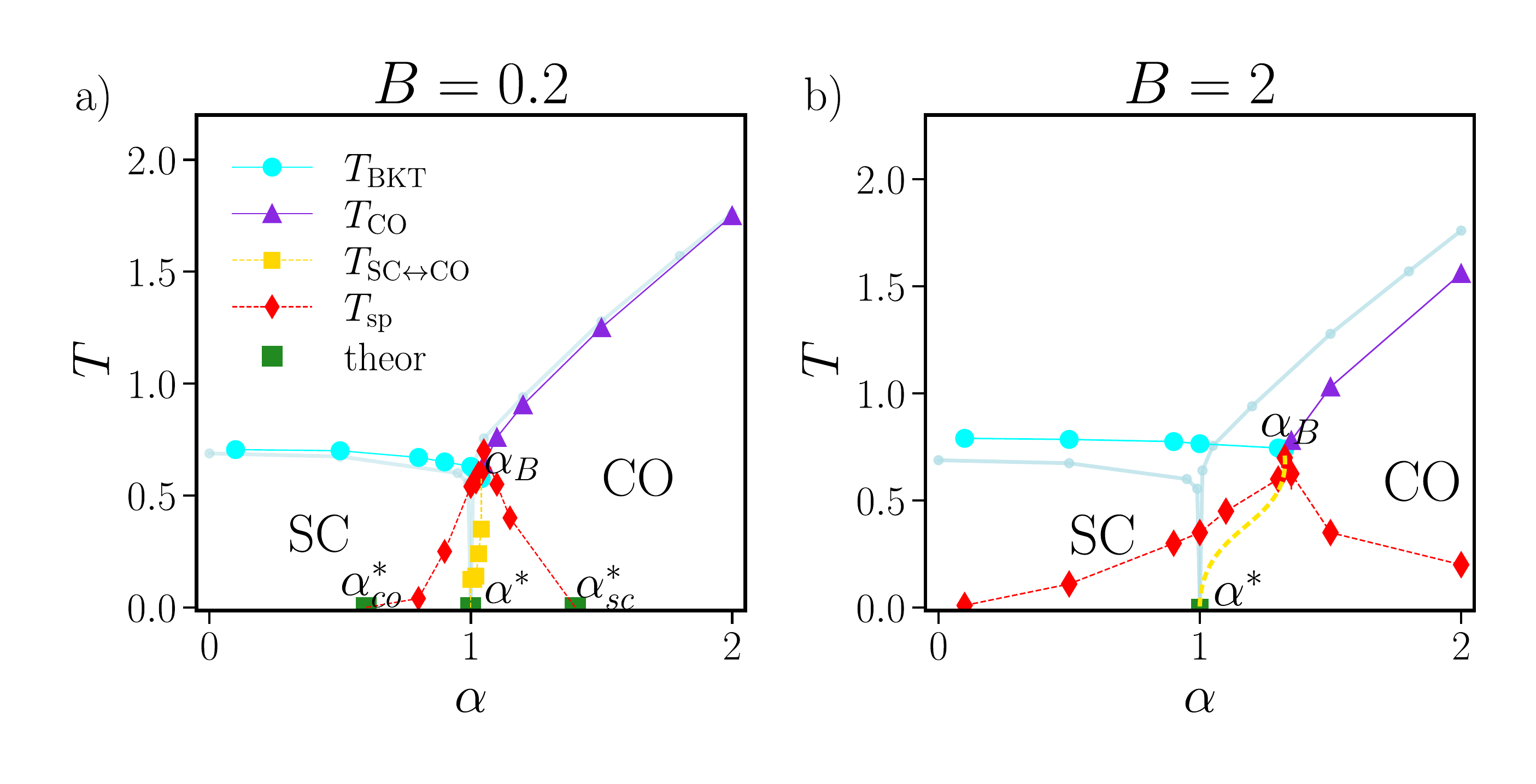}
    \caption{Phase diagram in the $T$\,vs.\,$\a$ plane for the competition between SC 
    and CO, modelled with 
    an XXZ model with a barrier height (a) $B=0.2$ and (b) $B=2$. The light blue 
    lines refer to the bare XXZ model, for comparison. Cyan dots are the $T_\text{BKT}$ points 
    calculated with the BKT scaling law [Eq.\,(\ref{eq:XXZ_scalingBKT})]; purple triangles refer to the 
    Ising (CO) 
    transition, for which the $T_\text{CO}$ points are found by use of the Binder cumulant $U_N$ [Eq.\,\eqref{eq:UN}]; 
    yellow squares locate the first-order transition, while red diamonds indicate the spinodal 
    points; green points at $T=0$ are calculated analytically; yellow squares and red 
    diamonds in (a) are computed from the effective free energies $F(m_z)$ and the locations of its minima
    $F_\text{min}(T)$, while in (b) they are inferred within the protocol described at the end of Sec.\,\ref{sec:meta} }
    \label{fig:phasediag_MC}
\end{figure}

The resulting phase diagram is reported in Fig.\,\ref{fig:phasediag_MC}a.
For comparison, we also report the phase diagram for $B=0$ (light-blue line).
Cyan circles and purple triangles refer, respectively, to the critical temperatures $T_\text{BKT}$ and $T_{\text{CO}}$ 
calculated using the scaling laws of $J_\text{s}$ [Eq.\,\eqref{eq:XXZ_scalingBKT}] and $U_N$ [Eq.\,\eqref{eq:UN}].
The points in green along the line $T=0$ are the analytical results: the two squares 
at $\a_\text{CO}^*=0.6$  and $\a_\text{SC}^*=1.4$ are the spinodal points, calculated as described in 
Sec.\,\ref{sec:barrier} [see Eq.\,\eqref{eq:a*}], and the green square at $\a^*=1$ is 
the value at which the free energy has three equivalent minima at $m_z=0,\,\pm 1$ (first-order phase transition at $T=0$). 

As anticipated, the presence of the barrier with $B=0.2$ shifts the superconducting transition line to 
higher temperatures, up to a value $\a_B=1.04$ of the anisotropy control parameter. For  slightly larger values, we recover the CO
(Ising) transition line, which is shifted downwards with respect to the case $B=0$. The bicritical point 
is shifted to $\a_B>1$ and the first-order-line between the charge-ordered and the superconducting state 
has a positive slope, indicating that entropy slightly favours SC. 

\begin{figure}
    \centering
        \includegraphics[width=0.9\linewidth]{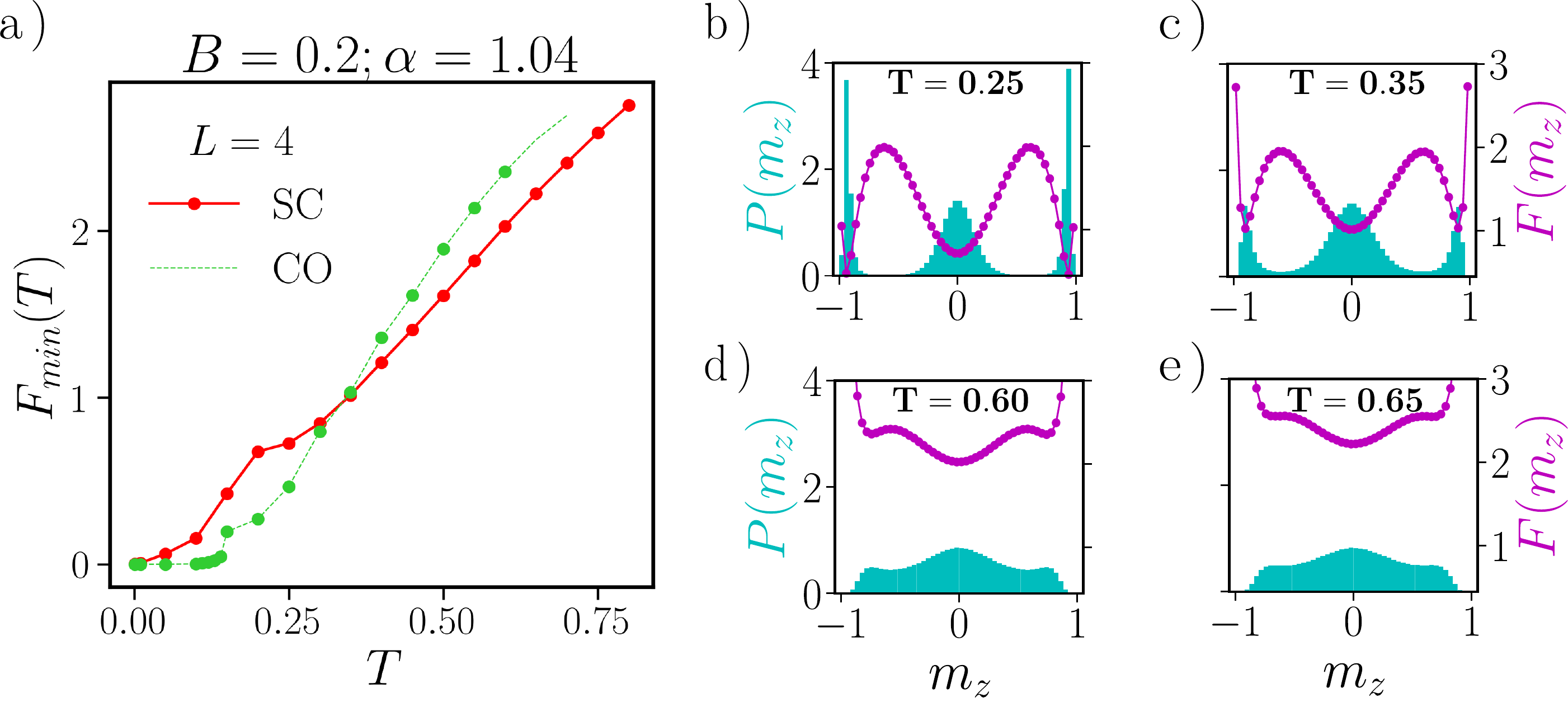}
    \caption{Effective free energies and probability distributions of $m_z$ for a system of linear size $L=4$ with 
    $\a=1.04$, $b=0.1$.
        (a) Local minimum of the free energy as a function of temperature $F_{\text{min}}(T)$. The red curve corresponds 
        to the minimum $F(m_z\approx 0)$  while the green line corresponds to the minimum $F(m_z\approx1)$. The crossing 
        temperature between the two lines marks the first order transition $T_{SC\leftrightarrow CO}$.
        (b-e) Effective probability density function $P(m_z)$ (turquoise) and free energy $F(m_z)$ (magenta) for:
        (b) $T=0.25$, 
        (c) $T=T_{SC\leftrightarrow CO}=0.35$.
        (d) $T=0.60$, and (e) $T=T_\text{sp}=0.65$.
        The free energies $F(m_z)$ at each temperature were constructed from the distribution of $m_z$ within 
        $N_\mathrm{MC}=5\times 10^5$, $\tau_\mathrm{MC}=5\times 10^5$ and $\tau_\mathrm{MC}=50$.}
    \label{fig:XXZ+Hb_phasediag_b0.1}
\end{figure}

The first-order transition line ($T_{SC\leftrightarrow CO}$, marked by the yellow squares) and the spinodal lines 
($T_\text{sp}$ red diamonds) in the phase diagram are obtained by constructing the effective distribution function 
$P_\text{eff}(m_z)$, as discussed in Sec.\,\ref{sec:barrier}.
We report in Fig.\,\ref{fig:XXZ+Hb_phasediag_b0.1}a the minima of the free energy
$F_\text{min}(T)$ as a function of the temperature, for the case $\a=1.04$, where the superconducting state 
is marked in red and the charge-ordered state in green. The crossing point between 
the two curves is the first-order critical temperature $T_{SC\leftrightarrow CO}$. We see that, in agreement with our previous discussion, CO is the stable phase at low temperatures, then, with rising the temperature, the system switches to superconductivity and then reaches the disordered state. We predict that in a very clean system close to the $p_{O(3)}$ point this phenomenon of superconductivity stabilized by temperature could be seen.

In panels b-e of Fig.\,\ref{fig:XXZ+Hb_phasediag_b0.1}, we report the histograms at the temperatures 
$T=$ 0.25, 0.35, 0.60, 0.65, where the distribution of $m_z$ is in turquoise (left axis) 
and the corresponding free energy $F(m_z)$ is in magenta (right axis). 
At $T=0.25$, $F(m_z)$ displays three minima, Fig.\,\ref{fig:XXZ+Hb_phasediag_b0.1}b, the global ones being at $m_z=\pm 1$ 
(corresponding to CO). By increasing the temperature, at $T=0.35$, the three minima 
become equivalent, Fig.\,\ref{fig:XXZ+Hb_phasediag_b0.1}c, while at $T\geq 0.35$ the global minimum is at $m_z=0$ 
(corresponding to SC). To define the spinodal temperature, at each $\a$, we fitted 
the data $F(m_z)$ in the region around $m_z=0.5$, and looked for the temperature 
at which the free-energy curvature changes from downward to upward. It can be 
seen, by comparing panels (d) and (e) in Fig.\,\ref{fig:XXZ+Hb_phasediag_b0.1} how the two 
minima at $m_z=\pm 1$ disappear when the temperature is increased from $T=0.60$ 
(panel d) to $T=0.65$ (panel e), where the curvature near $m_z=\pm 1$ appears to 
be flat.

\subsection{Phase diagram for $B=2$}
\label{b1}

Let us consider now a barrier parameter $B=2$.
The phase diagram in {Fig.\,\ref{fig:phasediag_MC}b} 
shows that the BKT line survives for values $\a\gg 1$, up 
to $\a_\text{B}=1.325$. 
The two spinodal points at $T=0$, calculated according to Eq.\,\eqref{eq:a*}, are located at 
$\a_\text{CO}^*=-3$ which correspond to a reversed interaction in the charge sector and $\a_\text{SC}^*=5$ (both outside 
the displayed range).
Again, the BKT and CO points are extracted as discussed in 
Sec.\,\ref{b01}: the cyan circles represent the $T_\text{BKT}$ 
temperature, computed with the scaling relation of the superfluid 
stiffness $J_\text{s}$ [Eq.\,(\ref{eq:XXZ_scalingBKT})]; and the purple triangles are used to mark the Ising 
transition temperature $T_\text{CO}$, computed from the finite-size scaling 
analysis of the Binder cumulant $U_N$ [Eq.\,\eqref{eq:UN}].
To determine the spinodal points at finite temperature (red diamonds), we rely on the protocol discussed 
at the end of Sec.\,\ref{sec:barrier}. 
{The yellow dashed line is 
a guide to the eye, to sketch the expected first-order transition line connecting
the $T=0$ transition point at $\alpha^*$ (green square) to the bicritical point
at $\alpha_\text{B}$.}

In Fig.\,\ref{fig:Nmsq_b0.1} we plot the superconducting ($\widetilde\chi^{xy}$) 
and charge-ordered ($\widetilde\chi^z$) mean-square magnetization (panels d and e, respectively), and the susceptibility
$\chi^z$ (panel f), at different values of the anisotropy parameter, in the range 
$0.1<\a<2$, for $B=2$. 
As one can see, $\widetilde\chi^{xy}$ is significant for values of anisotropy $\a\leq 1.325$, well above the 
isotropic Heisenberg limit.
The situation gets reversed as soon as $\a\geq 1.35$, where the main response of the system is in the out-of-plane 
direction (corresponding to CO).
{As in Sec.\,~\ref{b01}, the susceptibility $\chi^z$ shows precursor peaks of the charge-ordered state 
found at $\a\leq 1.35$, down to $\a=1.3$, where a very broad peak can observe.}

The case $\a=1.35$ highlights again the possibility of having a superconducting state stabilized by entropic effects. Indeed, 
upon cooling, the superconducting mean-square order parameter, light blue curve in Fig.\,\ref{fig:Nmsq_b0.1}c, follows the BKT behaviour lowering $T$,
similarly to the curve at $\alpha=1.325$ (green curve), down to $T=0.8$.
{By further lowering $T$ to $T=0.775$, $\widetilde{\chi}^{xy}$ drops down by a factor $\sim$500 and, correspondingly, $\widetilde\chi^{z}$ is increased by a factor $\sim 800$.}
It is worth noting how the peak in $\chi^z$ for this value of anisotropy is still very smeared as for $\a<1.35$, 
thus leaving no doubt about the nature of the ground state.

\begin{figure}
    \centering
    \includegraphics[width=0.8\linewidth]{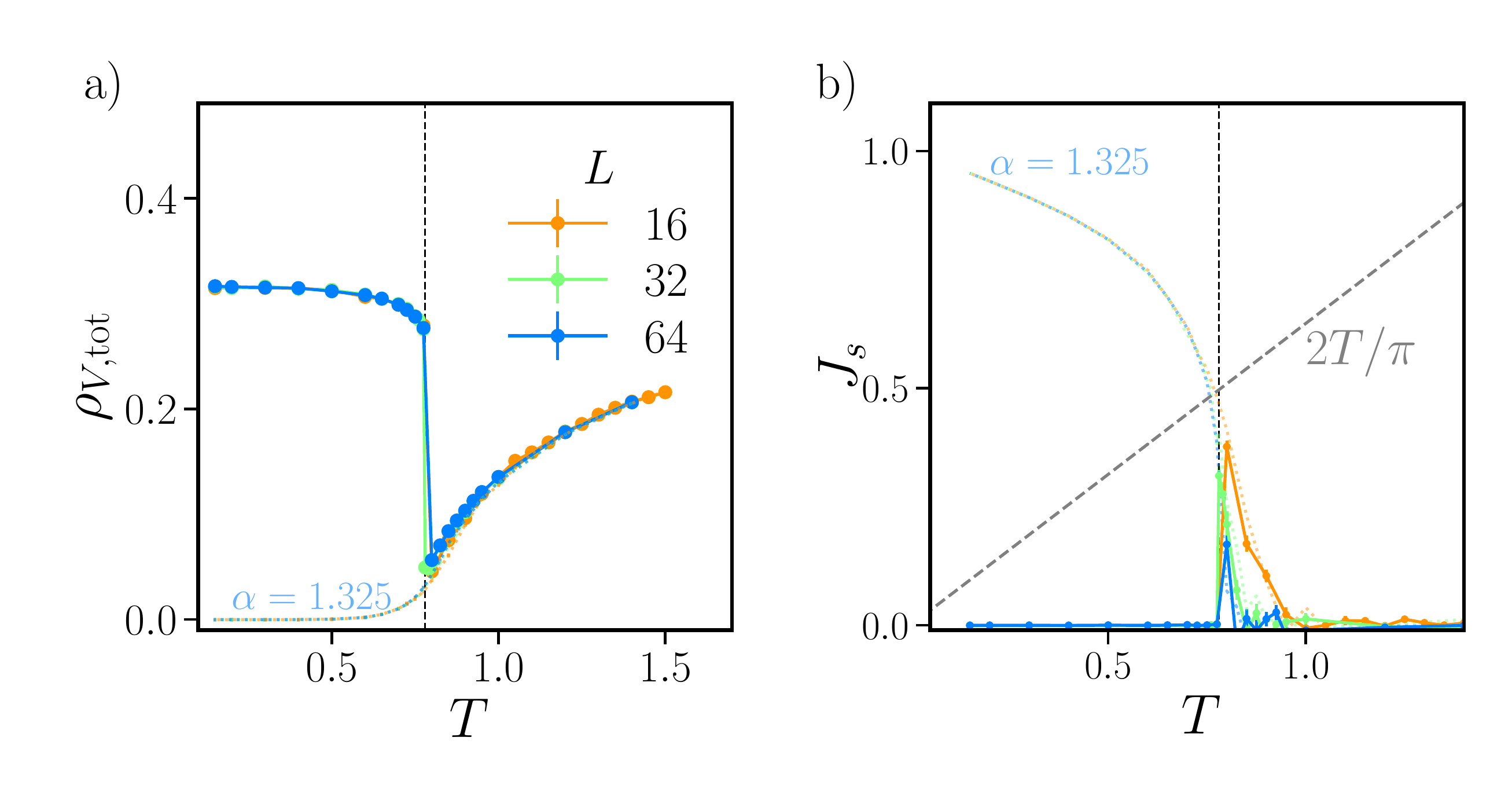}
    \caption{(a) Total density of vortices and antivortices $\rho_\text{V,tot}$ [Eq.\,\eqref{eq:vorticity}] as a function of the temperature, for $\a=1.35,\,B=2$, shows a re-entrant phase as $\rho_\text{V, tot}$ seems to decay exponentially lowering $T$ down to a temperature $T_\text{c}$ where vortices suddenly proliferates. $T_\text{c}$, marked with the vertical dashed line, was deduced from the Binder cumulant $U_N$. (b) The same trend is also confirmed by the finite-size effects in $J_\text{s}$. Error bars are calculated using the bootstrap resampling method with 100 datasets and blocks of size 100. The case $\alpha=1.325$, showing typical BKT feature, is plotted in lighter colours as a benchmark.}
    \label{fig:nV_b1}
\end{figure}

To fully describe the properties of the anomalous transition found at $\a=1.35$, 
we looked at the evolution in temperature of the total density of vortices, given in Eq.\,\eqref{eq:vorticity}. 
The $T$-dependence of $\rho_\text{V, tot}$, is shown in Fig.\,\ref{fig:nV_b1}a. 
In the BKT scenario, $\rho_\text{V,tot}$ is supposed to be exponentially 
suppressed as the temperature is lowered towards $T_\text{BKT}$, { as a 
consequence of the binding of vortex-antivortex pairs}.
This happens up to $\alpha=1.325$ (show as a benchmark with light colours in Fig.\,\ref{fig:nV_b1}a), where the 
suppression of 
$\rho_\text{V, tot}$ coincide with the appearance of a finite $J_\text{s}$ (light colours in Fig.\,\ref{fig:nV_b1}b).
The $\a=1.35$ curve seems to follow this trend in the high-temperature regime.  
Crossing the temperature $T\approx 0.8$, a sudden proliferation of free vortices is observed. This indicates an anomalous 
transition from an almost BKT-like superconducting state at high temperatures, turning into a charge-ordered state below 
$T\approx 0.8$, in agreement with the trend found in $\widetilde\chi^{xy}$ and $\widetilde\chi^z$. 
Note that such anomalous behaviour is also detected by finite-size effects in the superfluid stiffness 
plotted in  Fig.\,\ref{fig:nV_b1}b.
At high temperatures, the paramagnetic phase seems to be on the verge of undergoing a BKT transition, as it is visible 
from the tails of $J_\text{s}$, while instead at $T\simeq 0.76$ (vertical dashed line) the system 
develops CO and $J_\text{s}$ drops to zero.

\begin{figure}
    \centering
    \includegraphics[width=0.9\linewidth]{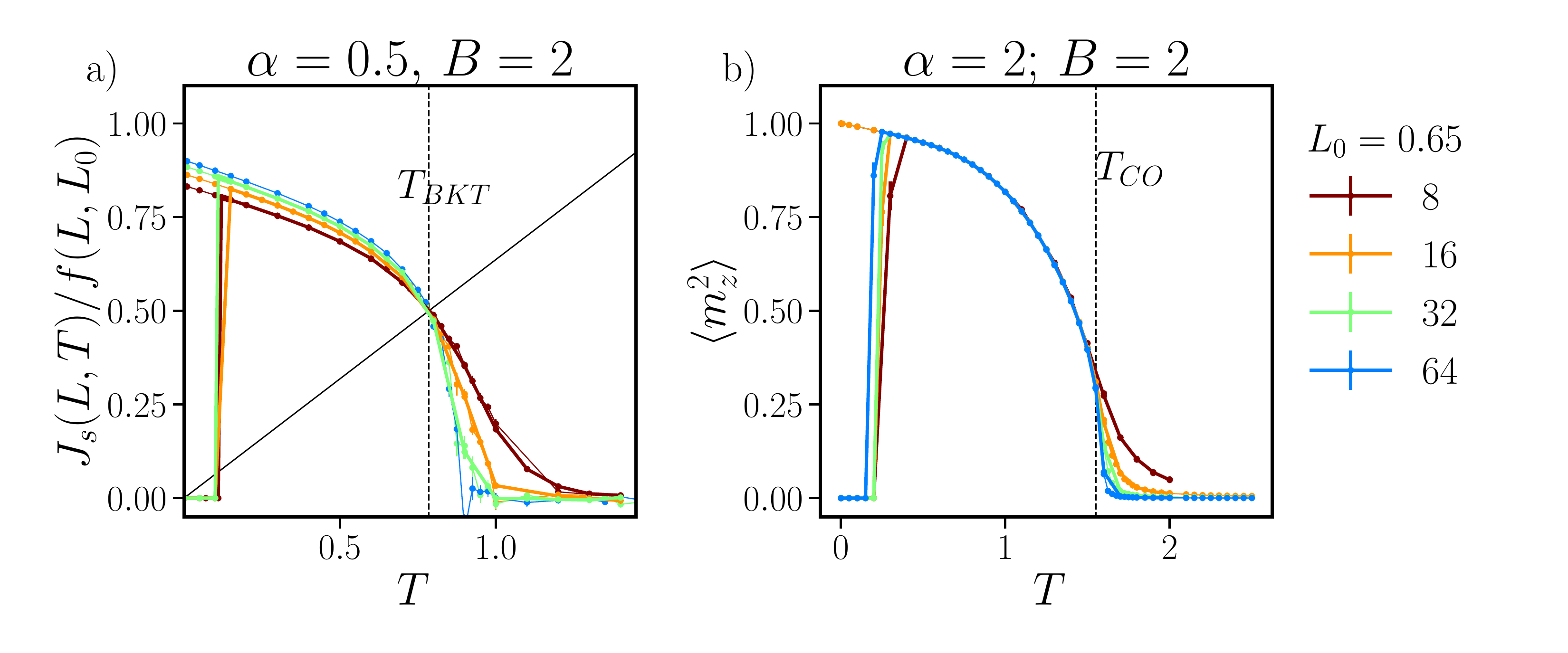}
    \caption{(a) Superfluid stiffness rescaled according to 
    Eq.\,(\ref{eq:XXZ_scalingBKT}) (in the label of the vertical
    axis we defined $f(L,L_0)=1+\left[2\ln(L/L_0)\right]^{-1}$ for brevity) for $\a=0.5$ and (b) mean-square charge-ordered 
    magnetization $\<m_z^2\>$ for $\a=1.5$ at various $L$ (color code as indicated in the legend). Thin lines correspond 
    to the usual cooling down protocol; thick lines stand for the results obtained when the system is heated up starting 
    from the metastable state -- (a) up, (b) parallel and on the $xy$ plane. Note that the temperature at which the 
    system jumps from the local to the global minimum state is not strongly dependent on the system size $L$. The error 
    bars are calculated using the bootstrap resampling method with 100 dataset and blocks of size 100.}
    \label{fig:metastable_b1}
\end{figure}

Finally, let us discuss in more detail the metastable states for $B=2$. 
As an example, we report in Fig.\,\ref{fig:metastable_b1} the superfluid 
stiffness, rescaled according to Eq.\,(\ref{eq:XXZ_scalingBKT}), for $\a=0.5$ (panel a) 
and the mean-square magnetization $\<m_z^2\>$  signalling CO, for $\a=2$ (panel b). The dots shown 
in the plot are computed cooling down the system from a random configuration at 
a given size $L$ while thick lines stand for the heating up process from the 
metastable state.
As one can see, the jump from the metastable state at low temperature to the ground state
is not strongly dependent on the system size $L$.

\section{Dirty system}
\label{dirty}

\begin{figure}
    \centering
    \includegraphics[width=0.95\linewidth]{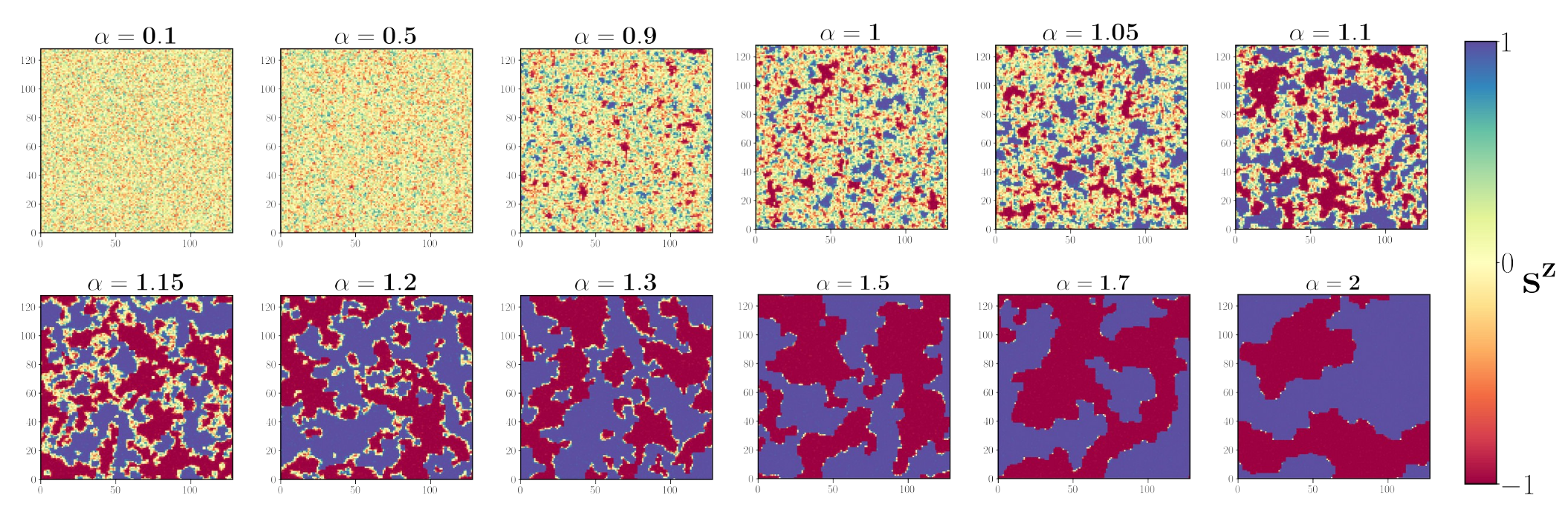}
    \caption{Snapshots of the final MC configuration at $T=0.001$ for systems of size $L=128$ and  $\a=0.1,\,0.5,\,0.9,\,1,\,1.05,\,1.1,\,1.15,\,1.2,\,1.3,\,1.5,\,1.7,\,2$. The colour code maps the $S^z$ component of the pseudospin, ranging from $+1$ (blue), to $0$ (in-plane, yellow), to $-1$ (red). }
    \label{fig:snapshots}
\end{figure}

We now discuss the role of disorder. The localizing effect of impurities is not
expected to significantly alter the BKT transition found at $\a<1$\cite{harris1974effect}.
For $\a>1$,  a study at zero temperature has shown that the effect of disorder is to break CO into a polycrystalline 
state \cite{attanasi2009competition,Leridon2020,caprara1995disorder}. This can be seen in 
Fig.~\ref{fig:snapshots}, where we show low-temperature snapshots ($T=0.001$) of the MC simulations for increasing $\a$.   
The colour code maps the CO order parameter ranging from $S^z=+1$ (blue), through 
$S^z=0$ (no CO, yellow), to $S^z=-1$ (red). We remind the reader that $S^z=\pm 1$ encodes two variants of the CO, e.g., 
with maxima of the charge density located at two different lattice positions, connected by translational symmetry of 
the lattice. It was argued in 
Refs.~\cite{attanasi2009competition,Leridon2020,caprara1995disorder} that at the boundary of such domains CO 
gets frustrated and FSC emerges. Indeed, 
as CO fluctuations are enhanced by increasing $\a$, 
the superconducting condensate gradually loses its two-dimensional 
nature by forming thinner and thinner filamentary structures. {As a consequence, as the superconducting cluster get narrower, a smearing of the BKT signatures is expected.}

To probe the BKT transition, we monitor the superfluid stiffness $J_s$ and its scaling according to 
Eq.\,\ref{eq:XXZ_scalingBKT}. 
However, this will not be sufficient in our discussion because of the gradual 
broadening of the BKT jump of $J_\text{s}$, along with the gradual violation  of the BKT scaling relation, Eq.\,(\ref{eq:XXZ_scalingBKT}). Moreover, a substantial fraction of in-plane pseudospins will survive also in the charge-ordered region of the phase diagram, as it is already visible from the snapshots in Fig.\,\ref{fig:snapshots}.
Thus, in the dirty system, we will also study the disorder-mediated superconducting correlation function among two
pseudospins separated by $\bm r$, defined as
\begin{equation}
    C^{xy}(\bm r)=\overline{\left\langle\sum_{\bm R} \left(S^x_{\bm R} S^x_{{\bm R}+{\bm r}} + S^y_{\bm R} S^y_{{\bm R}+
    {\bm r}}\right)\right\rangle}
       =\overline{\left\langle \sum_{\bm R}  \sin\vp_{\bm R} \sin\vp_{{\bm R}+{\bm r}} \cos(\theta_{\bm R} 
       - \theta_{{\bm R}+{\bm r}}) \right\rangle}.
\label{eq:C_bkt_XXZ}
\end{equation}
Note that the average over many disorder realizations (indicated by the overline) restores spatial isotropy at large distances.

Indeed, as it is well known, one of the hallmarks of the BKT topological phase transition is encoded in the 
peculiar behaviour of the correlation function:  
    \begin{align*}
    C^{xy}(\bm r)&\sim \mathrm e^{-|\bm r|/\xi^{xy}},\, \quad\quad T>T_\text{BKT},
        & \xi^{xy}=1/\ln(2T/J),\\
       C^{xy}(\bm r) &\sim \left(\frac{a}{|\bm r|}\right)^{\frac{T}{2\pi J}},\,\quad T\leq T_{BKT},
        & \xi^{xy}\rightarrow\infty,~~~~~~~~~~~~ \\
    \end{align*}
in the thermodynamic limit, where $J$ is the stiffness at $T=0$ and $a$ is the characteristic size of a vortex 
core \cite{patashinskii1979fluctuation}.
The infinite correlation length of superconducting fluctuations $\xi^{xy}$ at $T\leq T_{BKT}$ cannot be probed by numerical 
simulations of a finite system. Instead, in a MC simulation, one has $\xi^{xy}\propto L$.
It should also be noted that, as a consequence of the presence of out-of-plane fluctuations, 
Eq.\,(\ref{eq:C_bkt_XXZ}) acquires an extra factor $\sin\vp_{\bm R} \sin\vp_{{\bm R}+{\bm r}}$ 
with respect to the standard BKT case, in which 
$C^{xy}(\bm r)=\overline{\<\sum_{\bm R} \cos(\theta_{\bm R} - \theta_{{\bm R}+{\bm r}})\>}$.

The average in-plane component
\begin{equation*}
    \overline{\< \sin\varphi \>}=\overline{\langle \sum_\mathbf{R} \sin\varphi_\mathbf{R} \rangle}
\end{equation*}
provides instead a good estimate of the short-range SC still present in the system. It is worth noting that this 
quantity again does not contain the information about the coherence of the condensate, which is encoded in the 
$\theta$ variable. We checked, however, that for all values of $\a$ we investigated, at least at low temperatures, the 
spins belonging to the same superconducting cluster are indeed coherent.

Moreover, the charge-ordered state loses now its long-range order so that the resulting magnetization and the corresponding Binder cumulant in Eq.\,(\ref{eq:UN}), cannot be used to define $T_\text{CO}$. 
For $\a>1$, the system gradually evolves toward a random-field Ising model and the ground state appears as a rather inhomogeneous landscape, characterized by large charge-ordered
puddles.
Therefore, we will use the CO correlation function defined as
\[
{C^{zz}(\bm{r})}= \overline{ \left\langle\sum_{\bm R} S^z_{\bm R} S^z_{\bm R+\bm{r}}\right\rangle},
\]
which is expected to decay exponentially as $\sim \mathrm e^{-|\bm{r}|/\xi^z}$, to 
characterize the behaviour of the charge-ordered state, using $\xi^z$ as a fitting parameter.

The random-field Ising model in two dimensions has no finite critical 
temperature\cite{imry1975random} and is characterized by a finite 
low-temperature correlation length that grows exponentially with reducing 
the strength of the random field\cite{aharony1983low}. The presence of the barrier term in 
Eq.\,\eqref{eq:XXZ_Htot} further suppresses transverse pseudospin fluctuation at low temperatures and enhances  
the clustering of up/down charge-ordered regions, even at $\alpha\gtrsim 1$, 
thereby favouring the polycrystalline behaviour up to a finite temperature 
$T_\text{CO}$. 
\begin{figure}
    \centering
    \includegraphics[width=0.65\linewidth]{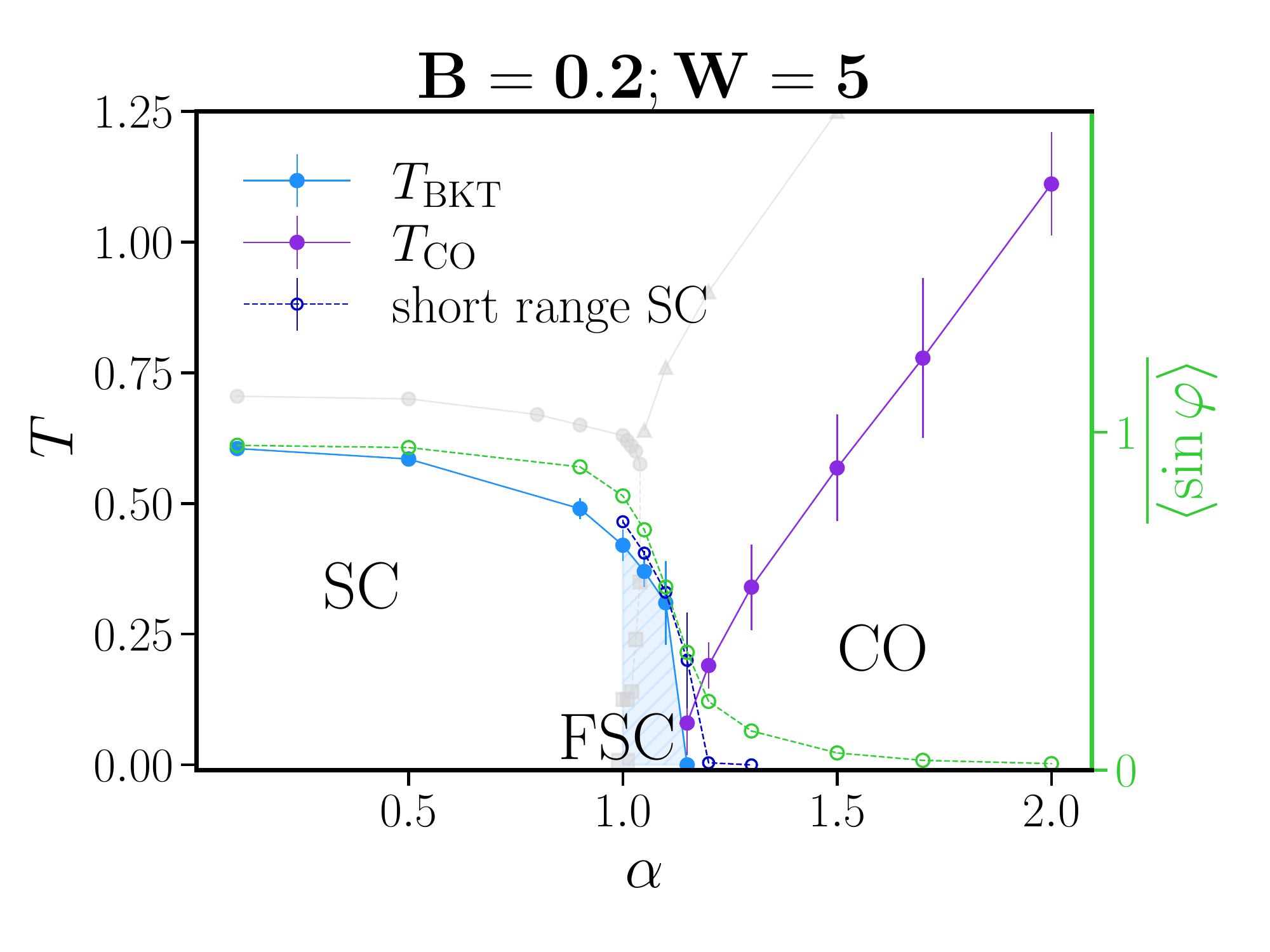}
    \caption{$T$\,vs.\,$\a$ phase diagram for $B=0.2$ and $W=5$ (light blue, purple and blue symbols and lines, left axis) 
    and average superconducting order parameter 
     $\overline{\<\sin\varphi\>}$ at at $T=0.001$ (green circles, right axis); the complete temperature dependence of 
     $\overline{\<\sin\varphi\>}$ can be found in Fig.\,\ref{fig:Js_a1.15}b. $T_\text{BKT}$ points (cyan) are computed using Eq.\,\ref{eq:XXZ_scalingBKT}; $T_\text{CO}$ points (purple) are computed from the linear fitting of $1/\xi^z$; short-range SC points (blue) refers to the temperature at which $J_s$ for $L=16$ crosses the critical line $2T/\pi$. The errorbars are calculated from the standard deviation of independent disorder configurations. The grey symbols show the phase 
    diagram of the clean system ($W=0$), for comparison.
    }
    \label{fig:phasediag_dirty}
\end{figure}

We estimate $T_\mathrm{CO}$ assuming the CO correlation length to behave 
as $\xi^z(T)\sim(T-T_\text{CO})^{-1}$, for $T$ approaching $T_\text{CO}$ from above (without getting too close 
to it), with the index
$\nu=1$ of the clean Ising model \cite{wu1966theory}. The idea is that, starting 
from high temperatures,
one can follow the critical behaviour of the clean Ising model, down to a temperature
at which the system crosses over to the non-critical behaviour of the random-field 
Ising model and the correlation length saturates to a finite value that 
determines the typical size of the clusters.

\subsection{Phase diagram for $B=0.2,\,W=5$}
\label{sec:dirty_phasediagram}

We present our results for barrier height $B=0.2$ and disorder strength $W=5$, to explore the effect of disorder in a 
situation when the first-order transition between the two phases would be nearly vertical in the clean case (see 
Sec.\,\ref{b01}). 

The phase diagram $T\,$vs$\,\a$ is reported in 
Fig.\,\ref{fig:phasediag_dirty}, where the $T_\text{BKT}$ points (cyan dots) 
are calculated using the BKT scaling 
law of $J_\text{s}$, while $T_\text{CO}$ (blue) is computed from the fit of $1/\xi^z$
in the temperature range where it exhibits a linear behaviour. 

For $\a<1$, the superconducting state is not much affected by disorder, except for a small 
suppression of the superfluid stiffness (see Appendix \ref{app:Js}).
Indeed, according to the Harris criterion\cite{harris1974effect}, the presence of 
spatially uncorrelated disorder does not alter the topological phase transition.

Instead, non-trivial features are expected for values $\alpha\gtrsim 1$ where spatially
correlated disorder emerges from the interplay between the competition and 
the presence of impurities. Our two striking results are indeed found on the 
$\a\geq 1$ side of 
the phase diagram: i) the observation of FSC 
for $\a\gtrsim 1$; and ii) the formation of a polycrystalline 
charge-ordered phase when $\a\gg1$. 

\begin{figure}[t]
    \centering
    \includegraphics[width=0.9\linewidth]{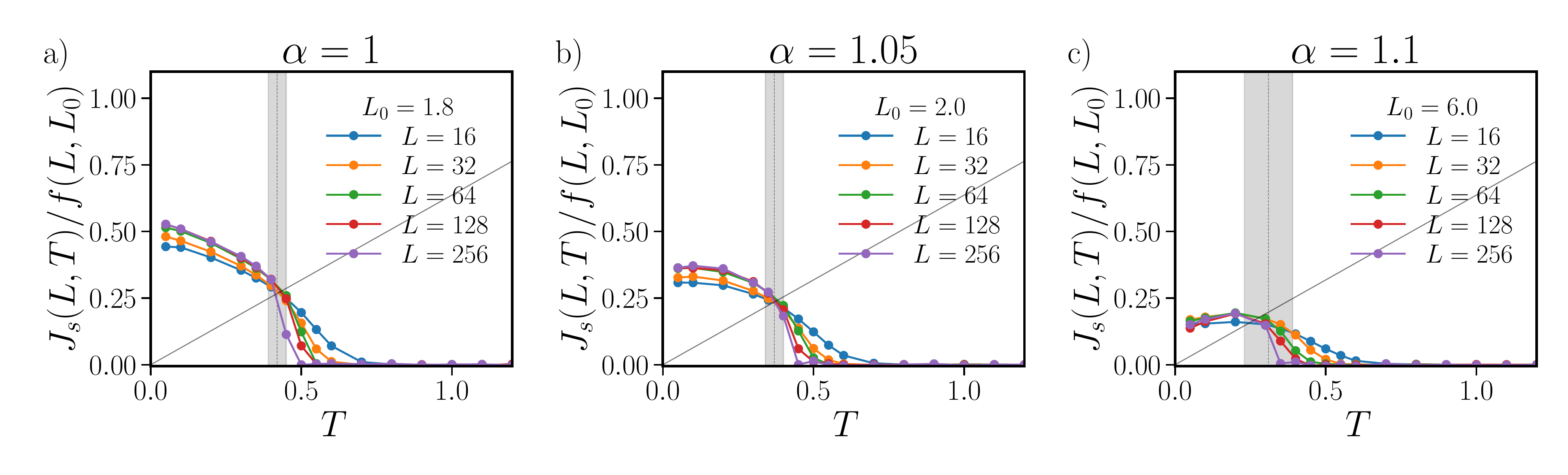}
    \caption{Crossing point of the superfluid stiffness rescaled according to 
    Eq.\,(\ref{eq:XXZ_scalingBKT}), with the BKT critical line $2T/\pi$ (full black line) at various linear sizes, 
    for $\a=1.1${, $B=0.2$, and $W=5$}. In the labels of the vertical
    axis we defined $f(L,L_0)=1+[2\ln(L/L_0)]^{-1}$ for brevity.}
    \label{fig:FSC_Js}
\end{figure}

{In Fig.\,\ref{fig:FSC_Js} we report the rescaled superfluid stiffness $J_\text{s}(L,T)$ at various $L$ 
(the grey line is the critical line $2T/\pi$), following the scaling law in Eq.\,\eqref{eq:XXZ_scalingBKT}. We find that the BKT scaling law works quite well up to $\a=1.1$, see Fig.\,\ref{fig:phasediag_dirty}), despite we observe deviations in the non-universal features of the phase transition.
The curves are obtained averaging over 
$N_{\text{dis}}=20$ disorder realizations for $L=16,32$, $N_{\text{dis}}=15$ for 
$L=64$, $N_{\text{dis}}=10$ for $L=128$, and $N_{\text{dis}}=7$ for 
$L=256$.
The vertical line and the grey shaded area correspond to $T_\text{BKT}\pm \sigma_{T_\text{BKT}}$ (the error being 
calculated as to include the smearing of the jump and the uncertainty on the fitting parameter $L_0$).

It is not surprising that
for $\a=1$ (panel a) we still observe a pretty clear jump of the stiffness, although the clustering of small 
charge-ordered
regions in the system can already be observed (see the corresponding snapshot in Fig.\,\ref{fig:snapshots}).
In fact, as we showed in Sec.\,\ref{b01}, in the clean system the potential barrier stabilizes the superconducting state 
up to $\alpha_B=1.04$ .

Going towards $\a=1.05$ (panel b) we can still observe a well-defined crossing 
of $J_\text{s}$ with the critical line, whereas in the clean system this value corresponded already to a 
charge-ordered global minimum of the free energy.}

Finally, for $\a=1.1$,
we still find a finite superfluid stiffness,
and yet the usual BKT scaling relation, Eq.\,(\ref{eq:XXZ_scalingBKT}), has noticeable deviations as one can realize by scrutiny of 
Fig.\,\ref{fig:FSC_Js}c.  Rescaling the curves according to Eq.\,(\ref{eq:XXZ_scalingBKT})
 leads to a spread of crossing point with the critical line 
$2 T/\pi$. Notwithstanding that, notice that compared with the unscaled curves shown in Fig.~\ref{fig:app3}c) of Appendix\,\ref{app:Js},  the different curves here have a convergence to a small region of temperatures showing approximate scaling. 
Using the BKT scenario, we obtain 
$T_\text{BKT}=0.31\pm 0.08$. This is consistent   with the (negative) minimum value of 
the derivative of the superfluid stiffness with respect to temperature (see Refs.\,\cite{lee2005helicity, lee2005monte} and Fig.~\ref{fig:app3}d) in Appendix \ref{app:Js}).
To take into account uncertainties in the definition of the BKT critical temperature, we considered a conservative confidence interval highlighted in gray in Fig.\,\ref{fig:phasediag_dirty}c. The difficulties in applying scaling relations in this case can 
be linked to the emergence of new length scales,
presumably related to the geometrical structure of the system, along with 
the typical sizes of vortex-antivortex pairs in the BKT theory.

It is very interesting that in this regime the stiffness decreases with temperature. This can be seen as a remnant of the entropy-induced superconductivity observed in the clean system, which disfavours superconductivity at low temperatures. Such an anomaly may be measured in samples close to the $p_{O(3)}$ point and is another prediction from this work. 

Note that the downward curvature in the 
low-temperature limit is not the consequence of finite-size effects, since 
a $256\times 256$ lattice with periodic boundary conditions provides 
already a reasonable size to observe reliable thermodynamic quantities. 

\begin{figure}[t]
    \centering
    \includegraphics[width=0.32\linewidth]{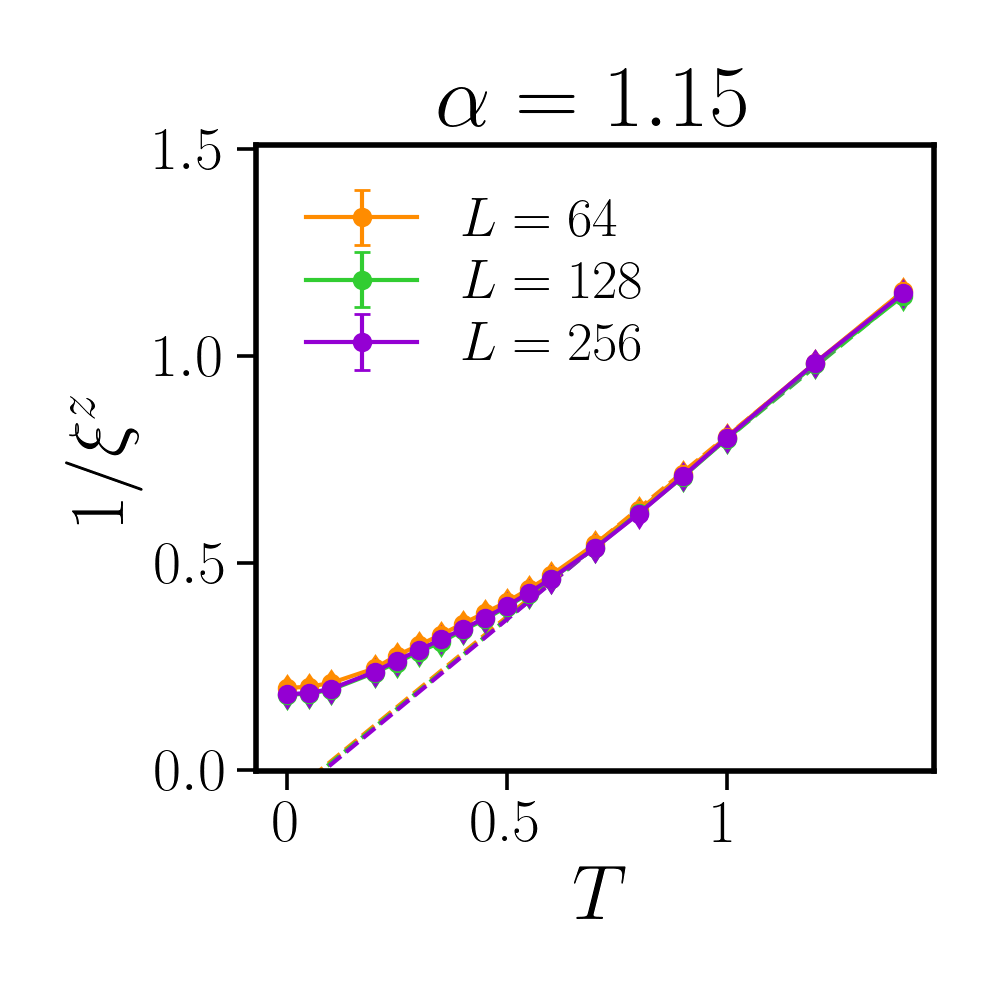}
    \includegraphics[width=0.32\linewidth]{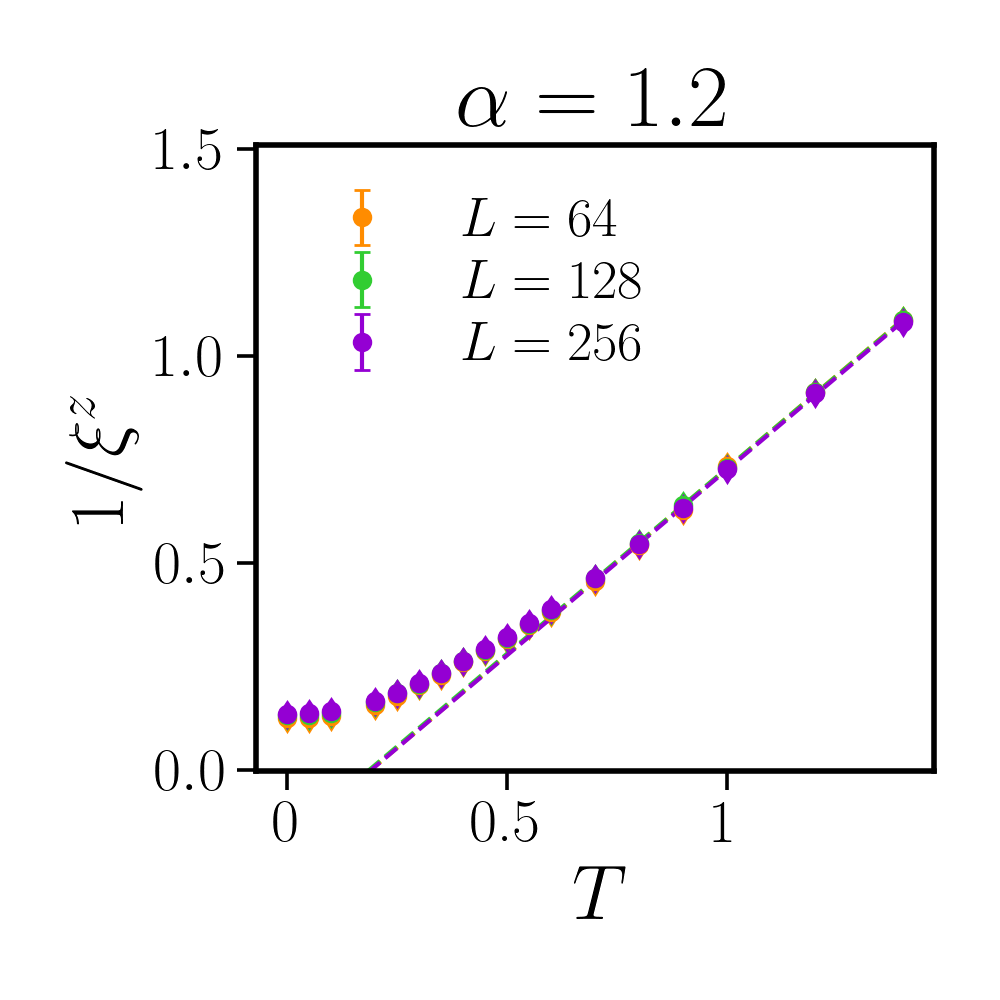}
    \includegraphics[width=0.32\linewidth]{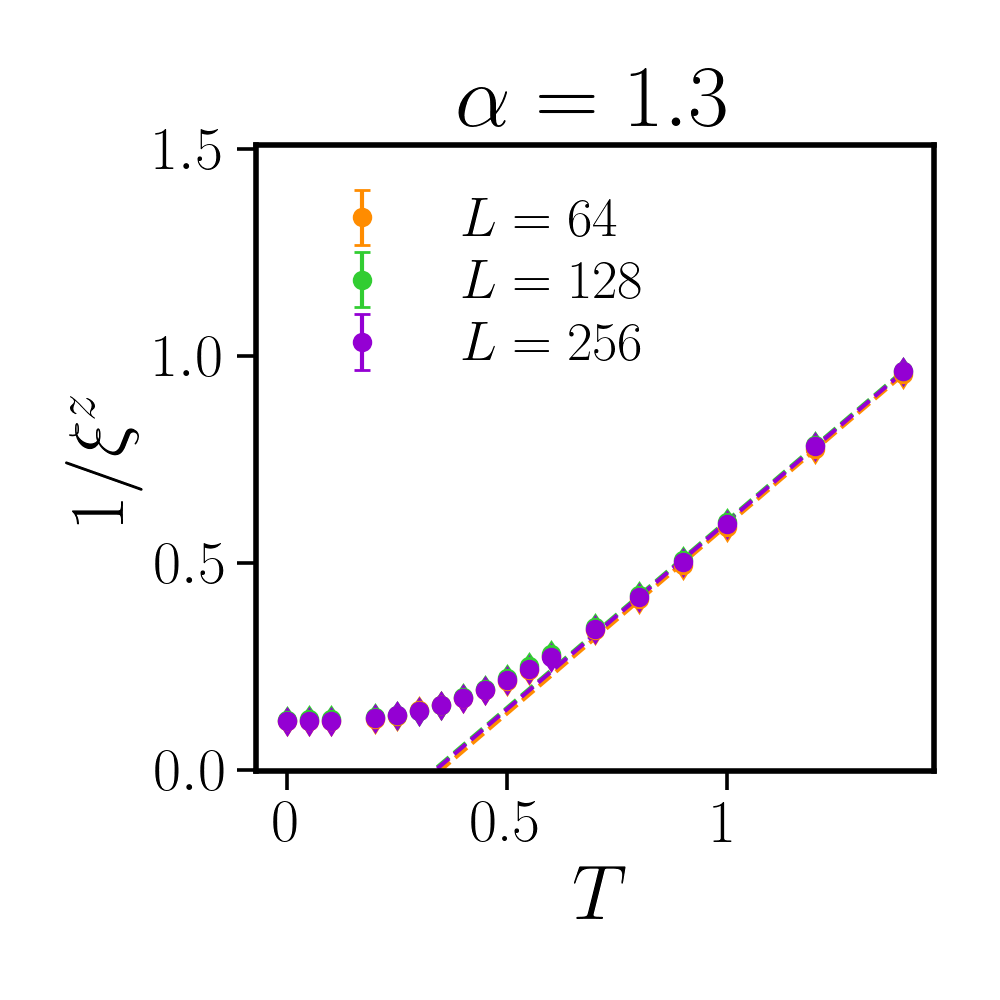}\\
    \includegraphics[width=0.32\linewidth]{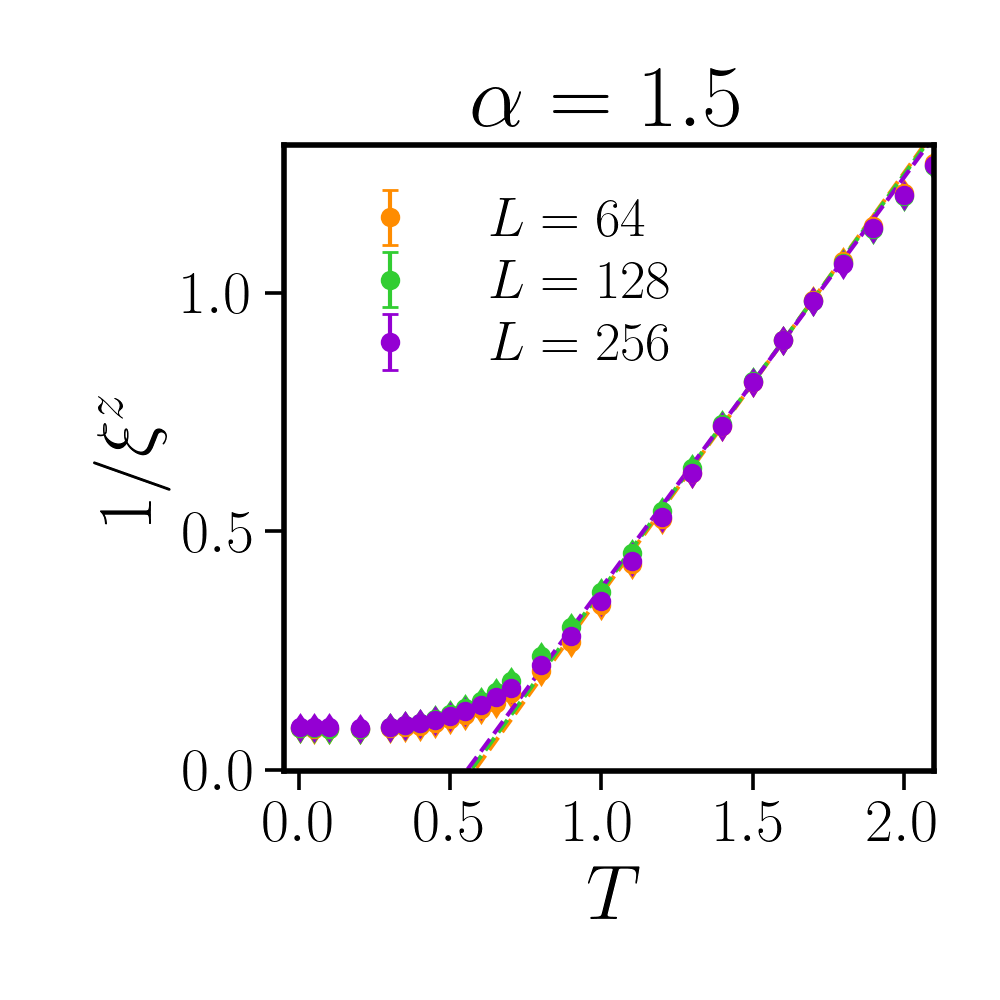}
    \includegraphics[width=0.32\linewidth]{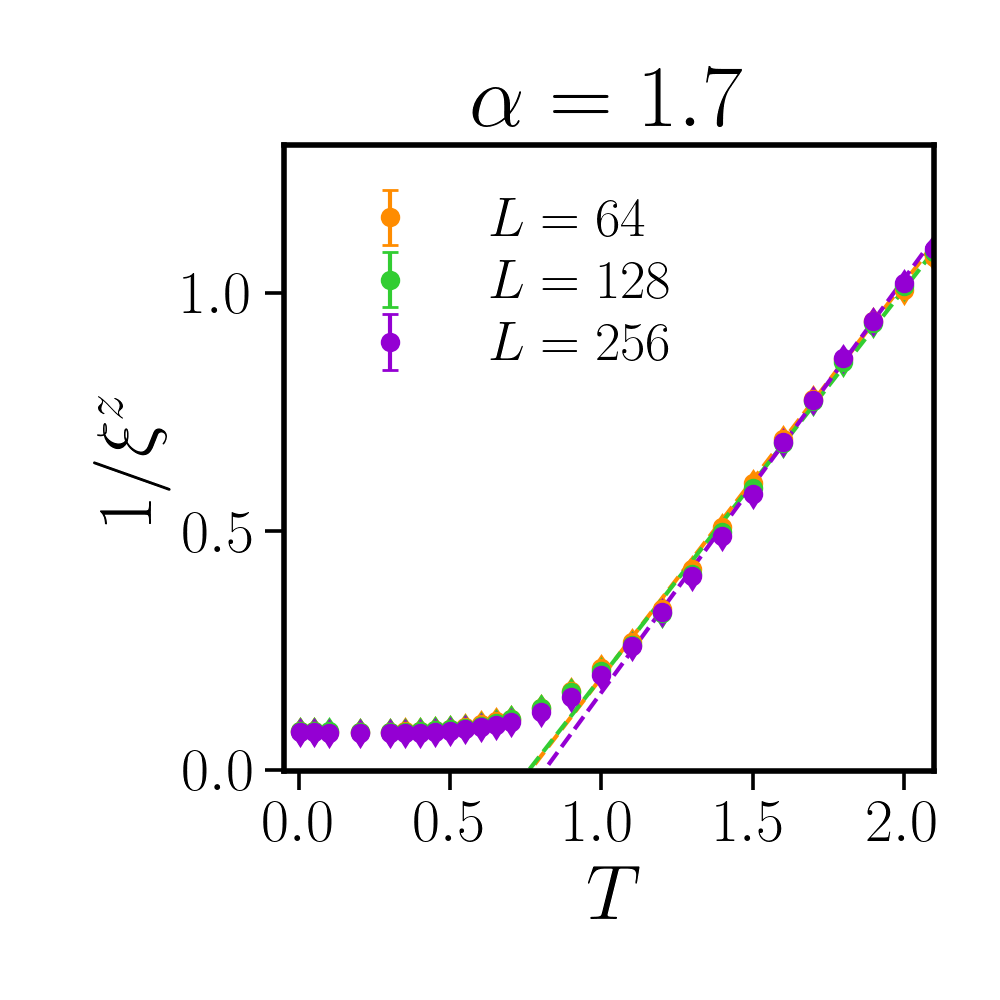}
    \includegraphics[width=0.32\linewidth]{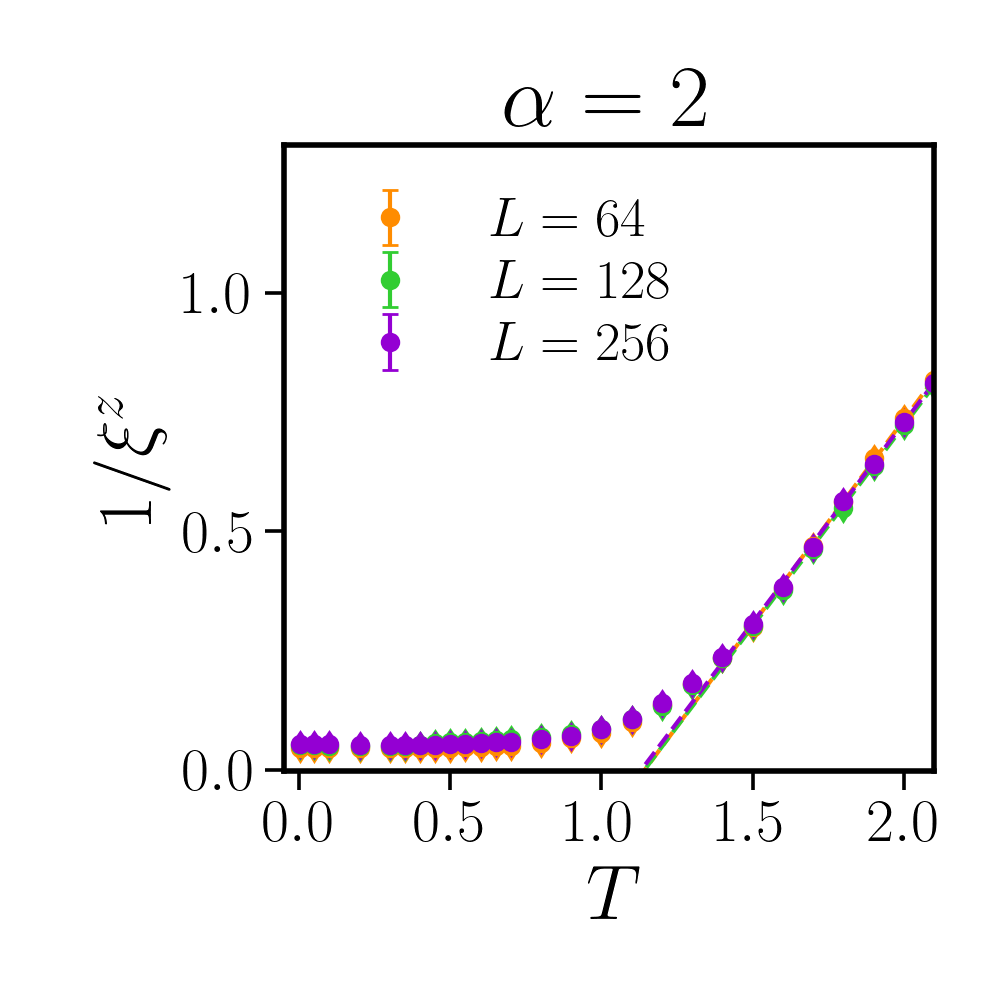}
    \caption{Inverse CO correlation length, $1/\xi^z(T)$, for {$B=0.2$,  $W=5$,} different values of the anisotropy $\a$, and sizes $L$. The dashed 
    lines are the linear fit of $1/\xi^z$.The errorbars are calculated from the standard deviation $N_\text{dis}$ of independent disorder configurations and $N_\text{dis}$=15, 10, 10 respectively for $L=$64, 128, 256.}
    \label{fig:xi_z}
\end{figure}

Let us discuss the CO correlation lengths $\xi^z$, used to define the polycrystalline charge-ordered phase 
for $\a\geq 1.15$ (see purple dots in Fig.\,\ref{fig:phasediag_dirty}).
The inverse correlation lengths $1/\xi^z$ as a function of temperature are displayed in Fig.\,\ref{fig:xi_z} for sizes 
$L=64,\,128,\,256$. The dashed lines correspond to the linear fits, and the different colours refer to the system 
size $L$, as shown in the legend.
As one can see, the linear decrease in temperature of $1/\xi^z$ deviates towards a constant 
plateau when the temperature is lowered below a finite value, the intercept of the linear fit defining the CO critical 
temperature $T_\text{CO}$. Moreover, all the curves show no 
sign of scaling and the low-temperature saturation value only depends 
on the anisotropy $\a$ (and on the barrier $B$ and on the strength of disorder $W$, if they were allowed to vary), 
thus signaling the presence of an intrinsic length scale related to the clusters. The downward deviation from linearity 
at high temperature, observed in the studied temperature interval when $\alpha\ge 1.5$, signals that the system is 
exiting the critical regime of the clean Ising model with further increasing $T$. {We would have observed the same deviation for $\a<1.5$ by looking at higher temperatures.}

We conclude this section by observing that the curves in Fig.\,\ref{fig:xi_z}
resemble the behaviour of the full width at half maximum of the CDW 
peak [proportional to $(\xi^z)^{-2}$] probed in YBa$_2$Cu$_3$O$_{7-\delta}$ and Nd$_{1+x}$Ba$_{2-x}$Cu$_3$O$_{7-\delta}$, 
by means of resonant inelastic X-ray scattering in Ref.\,\cite{arpaia2019dynamical}. There, the extrapolated $T_\text{CO}$ 
coincides with the 
temperature at which CO would occur once SC is suppressed by a magnetic 
field\cite{wu2011magnetic,Gerber2015}, while the saturation at low temperature, in the absence of a magnetic field, signals 
that CO competes with SC, and SC is more stable. Here, instead, the CO temperature obtained by this criterion near the $O(3)$ point is much lower than the asymptotic value at large $\alpha$. We will come back to this important point in the conclusions.

\begin{figure}
    \centering
    \includegraphics[width=\linewidth]{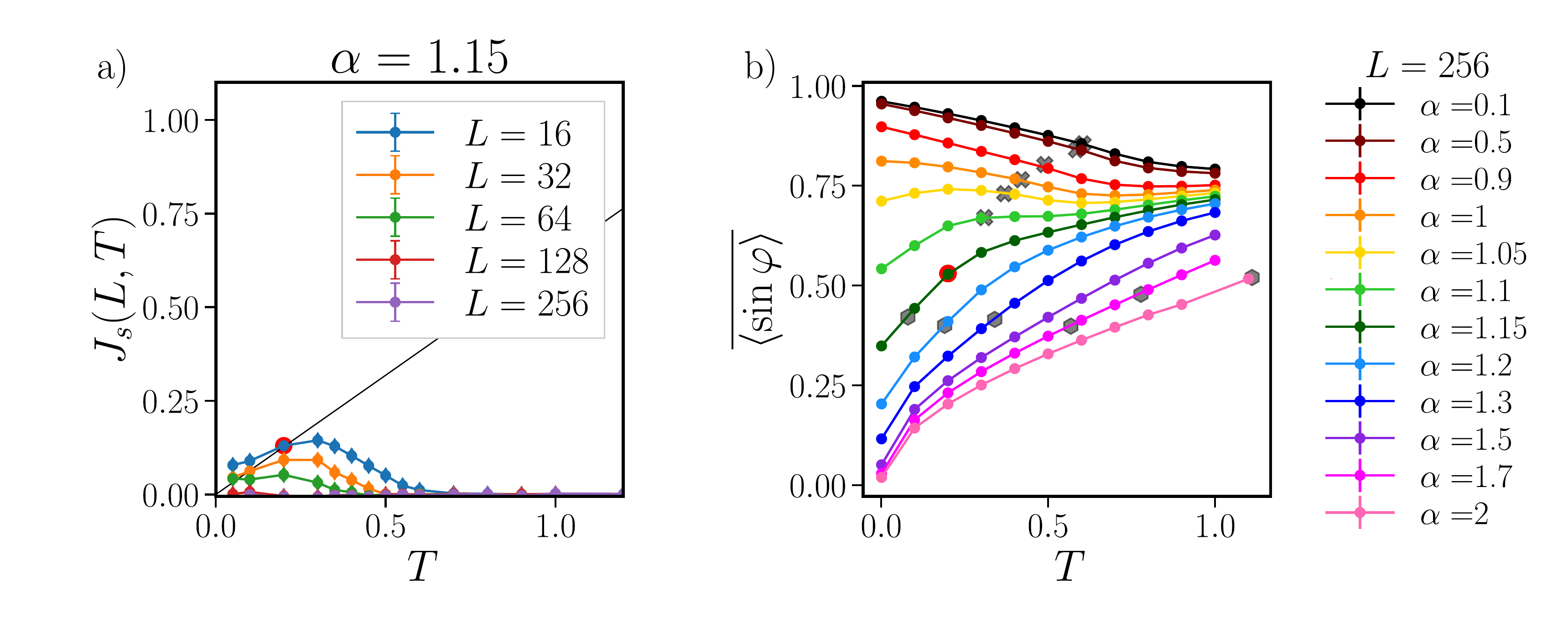}
    \caption{(a) Superfluid stiffness $J_\text{s}$ at $\a=1.15$, $B=0.2$, and $W=5$, for different system sizes. {MC parameters are $N_{\text{MC}}=2\times 
    10^3$, $\tau_\text{MC}=100$, $N_\text{out}=2\times 10^5$; $N_\text{dis}$ are 25 ($L=16$), 20 ($L=32,\, 64$) and 15 
    ($L=128,\,256$).} The errorbars are calculated from the standard deviation of $N_\text{dis}$ independent disorder configurations. (b) superconducting order parameter component $\overline{\< \sin\varphi\>}$ as a function of 
    temperature for various anisotropies $\alpha$ at {$L=256$}. Crosses are $T_\text{BKT}$ points and hexagons are the $T_\text{CO}$ points. The red dot highlights the short-range SC in both panels. }
    \label{fig:Js_a1.15}
\end{figure}

\subsection{Short-range and filamentary superconductivity}
\label{sec:short-rangeSC}

We discuss now the survival of a filamentary superconducting cluster and the presence of short-range SC in the 
polycrystalline charge-ordered side of the phase diagram.
Whereas in the case $\a=1.1$ it was still possible to define a BKT transition, albeit with a certain degree 
of uncertainty, when the anisotropy parameter is increased up to $\a\geq 1.15$ no $T_\text{BKT}$ can be defined from the 
crossing point. In particular, at $\a=1.15$ one can see from Fig.\,\ref{fig:Js_a1.15}(a) that $J_\text{s}$ vanishes 
already at $L=128$.
{The point $\a=1.15$ deserves, however, more attention since it displays both a finite critical 
temperature $T_{\text{CO}}$ and a short-range coherence of the superconducting cluster, as indicated by the finite value 
of $J_\text{s}$ at $L=16,\,32,\,64$.
The substantial fraction of superconducting pseudospins can also be observed by comparing the snapshots in 
Fig.\,\ref{fig:snapshots} (yellow component), in particular, those corresponding to $\a=1.1,\,1.15,\,1.2$.
Therefore, for $\a\geq 1.15$, despite the fact that the nearly one-dimensional nature of the 
superconducting cluster does not allow for the binding of 
vortex-antivortex pairs \cite{koma1995classical}, the finite residual 
superconducting component can still exhibit some short-ranged 
stiffness.
It is worth noting, once again, that in our coarse-grained model the
spacing of the pseudospin lattice $a'$ corresponds to the Josephson scale, i.e., $a'\approx\xi_\text{J}\approx 11a $ (see Sec.\,\ref{intro}), 
meaning that $L=16$ corresponds to about $(16\cdot 11)^2 \approx 31000$ quantum atoms. Such a large coherent region should have a strong impact in transport properties. While we have not computed the resistivity, it is clear that it will be quite small, as large patches of coherent regions will short-circuit the sample. We speculate that a broad transition should be observed with a large drop of the resistivity to a small but finite value.  
We thus include in our phase diagram the temperature at which we find short-range SC
(blue symbols) in the FSC region of our phase diagram of Fig.~\ref{fig:phasediag_dirty}. Those points indicate
the crossing of $J_\text{s}$ at $L = 16$ with the universal critical line $2T /\pi$.
A finite, even if exponentially small, stiffness is found up to values $\a=1.2$. This behaviour of $T_c\approx 0$ superconductivity is reminiscent of transport experiments in cuprates~\cite{shi2014two,Leridon2020}.

In order to get a more quantitative idea of short-range SC, in Fig.\,\ref{fig:Js_a1.15}(b) we show the average in-plane 
component
$\overline{\< \sin\varphi \>}$ as a function of temperature, for different values of $\alpha$. 
Note that in the standard BKT model, with purely planar pseudospins ($\alpha=0$), this should be identically equal 
to one at $T=0$. However, the presence of out-of-plane (corresponding to CO)
fluctuations renormalizes it to a lower value, that decreases with increasing $\alpha$. 
We indicate with gray crosses $T_\text{BKT}$ ($\a<1.15$) and with hexagons $T_\text{CO}$ ($\a\geq1.15$) discussed above 
and presented in the phase diagram (Fig.\,\ref{fig:phasediag_dirty}).
For $\alpha<1$, where SC is well described within the BKT scenario and no spatially correlated disorder 
emerges, $\overline{\< \sin\varphi \>}$ increases quite monotonically with lowering the temperature.
As $\a=1$ (in orange) we still observe the monotonic increase of $\overline{\< \sin\varphi \>}$ with decreasing $T$, and from superfluid stiffness computations we know that the system still exhibits quite clear BKT signatures (see Section \ref{sec:dirty_phasediagram} and Appendix \ref{app:Js}).

For $1<\a<1.15$, at high $T$ we observe first a slow decrease of $\overline{\< \sin\varphi \>}$ with increasing the temperature followed by an inflection point at $T_\text{infl}\gtrsim T_\text{BKT}$.
This range of the control parameter $\alpha$ lies inside the region of the phase diagram that we labelled 
with FSC in Fig.\,\ref{fig:phasediag_dirty}.
We stress again that up to $\a<1.1$, it is still possible to define the BKT temperature from the jump of the superfluid stiffness, which is smeared out but still clearly visible.
For $\alpha=1.1$ (light green), instead, the BKT scaling law starts showing deviations, and we observe a downturn 
of $\overline{\< \sin\varphi \>}$ at $T_\text{down}<T_\text{BKT}$. This may be related again to the entropically favoured SC of the clean case, which might also be the cause of the downturn of $J_s$ at very low temperatures. 

The curve for $\a=1.15$ (dark green) highlights again an interesting crossover scenario, which presents a filamentary pattern, clearly visible in the snapshots, but no long-range stiffness [see Fig.\,\ref{fig:Js_a1.15}a]. 
In this case, the decrease at high temperature follows a behaviour similar to the one found for $\a=1.1$, but with no inflection point, down to 
$T=0.2$ (indicated with a red dot). 
By further lowering the temperature, $\overline{\sin\varphi}$ becomes steeper.
The absence of an inflection point in $\overline{\sin\varphi}$ might be a proxy that the entropically favoured superconducting  state is now suppressed by the large CO fluctuations, although a small $J_\text{s}$ survives at finite $L$, becoming exponentially small with increasing the size. 
In fact, by comparing $\overline{\sin\varphi}$ for $\a=1.15$ with the corresponding $J_\text{s}$, one can observe that short-range SC is still present ($L<128$). 
Note that the curve for $L=16$ of Fig.\,\ref{fig:FSC_Js} has a maximum for $T=0.3$, then decreases by further lowering $T$, crossing the critical line $2T/\pi$ at T=0.2 (red dot). 
We point out that at the lowest temperature $T=0.001$ a substantial superconducting residue survives 
$\overline{\< \sin\varphi \>}=0.35$, presenting phase coherence. 
Even increasing the anisotropy up to $\a=1.2$, the superconducting fraction
is still about $20\%$.
We thus include $\overline{\< \sin\varphi \>}$ at $T=0.001$ and $L=256$ in our phase diagram in 
Fig.\,\ref{fig:phasediag_dirty} (right axis, in green) to stress out the presence of a macroscopic superconducting 
residue, that can show signatures in transport experiments 
even if it lacks long-range coherence.}

\begin{figure}
    \centering
    \includegraphics[width=0.24\textwidth]{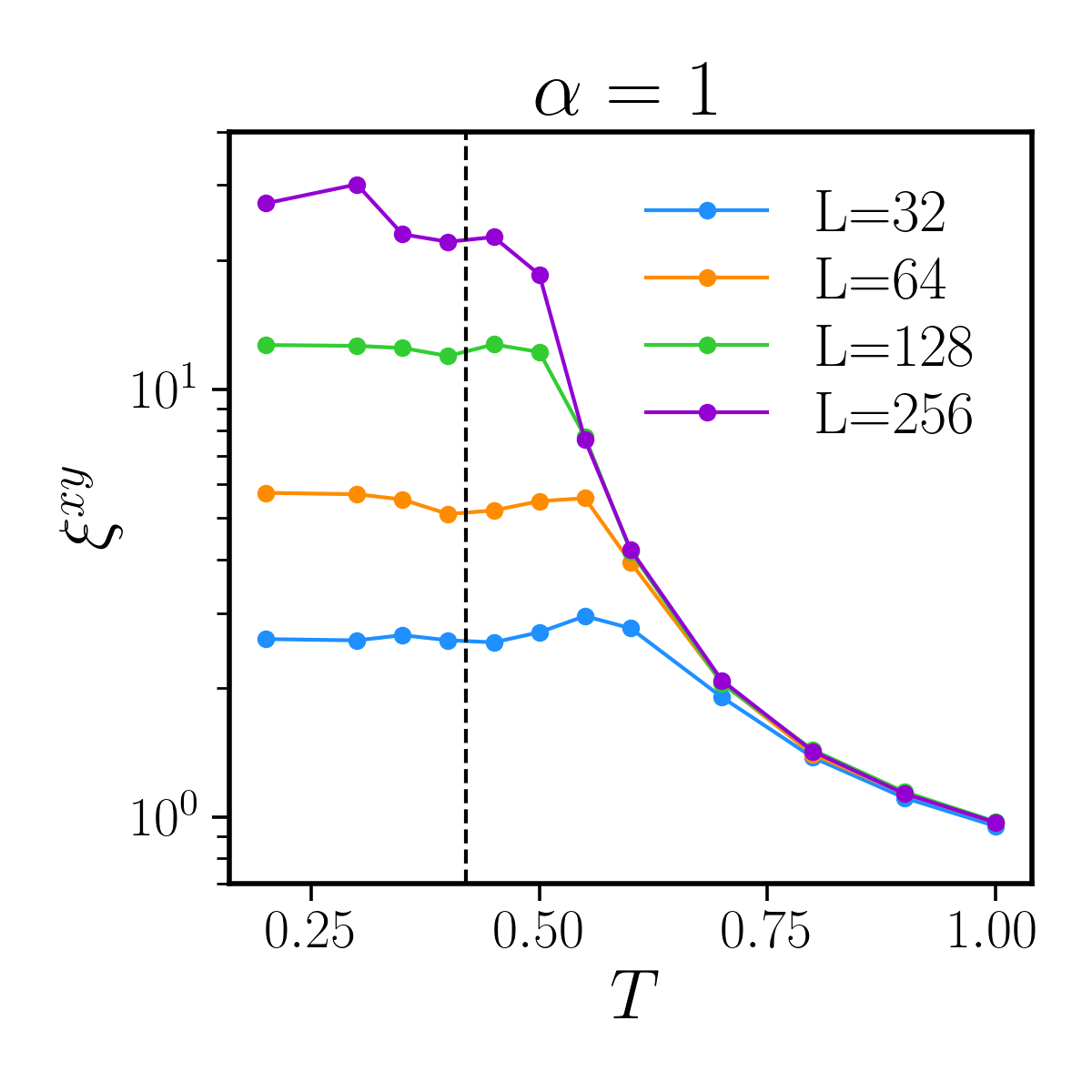}
    \includegraphics[width=0.24\textwidth]{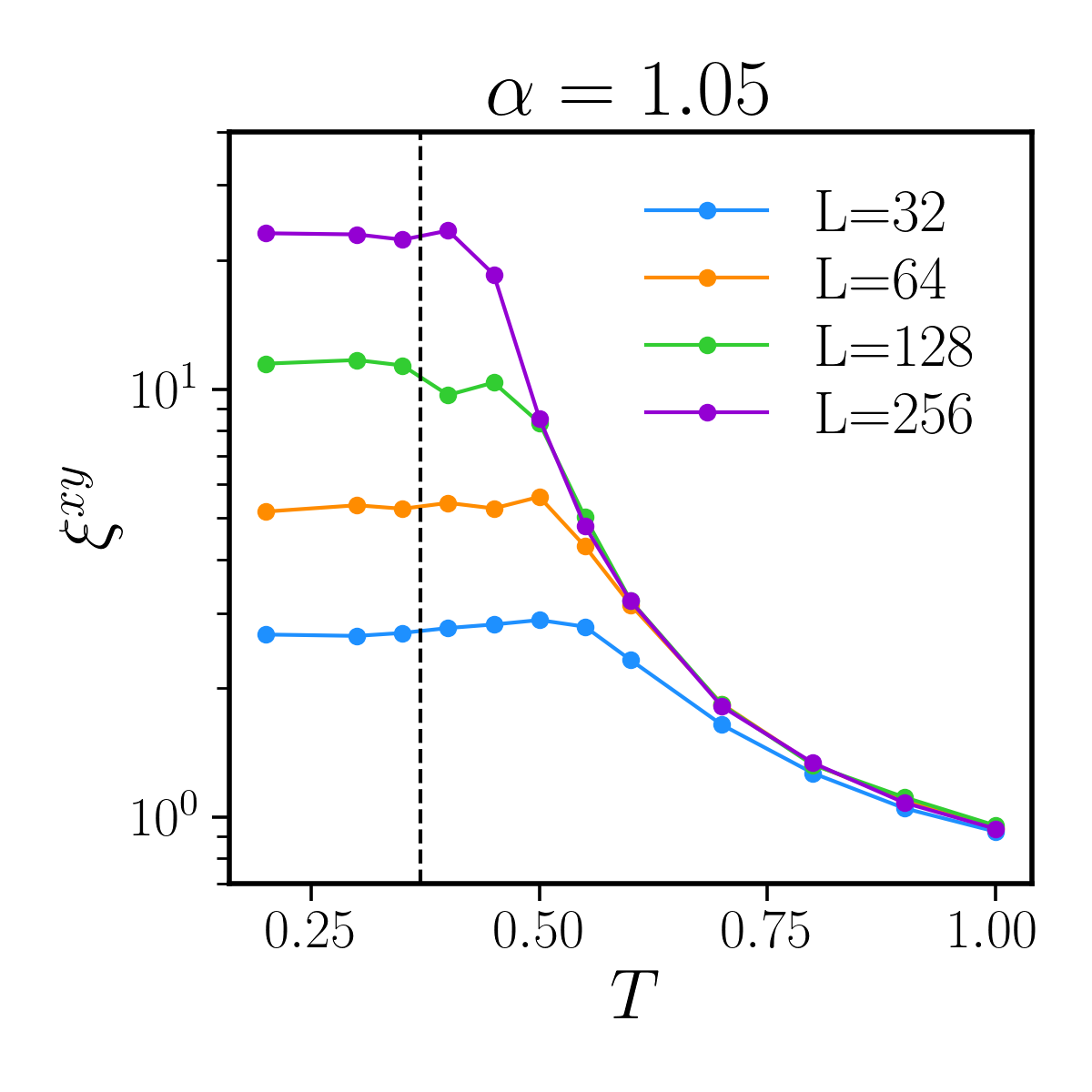}
    \includegraphics[width=0.24\textwidth]{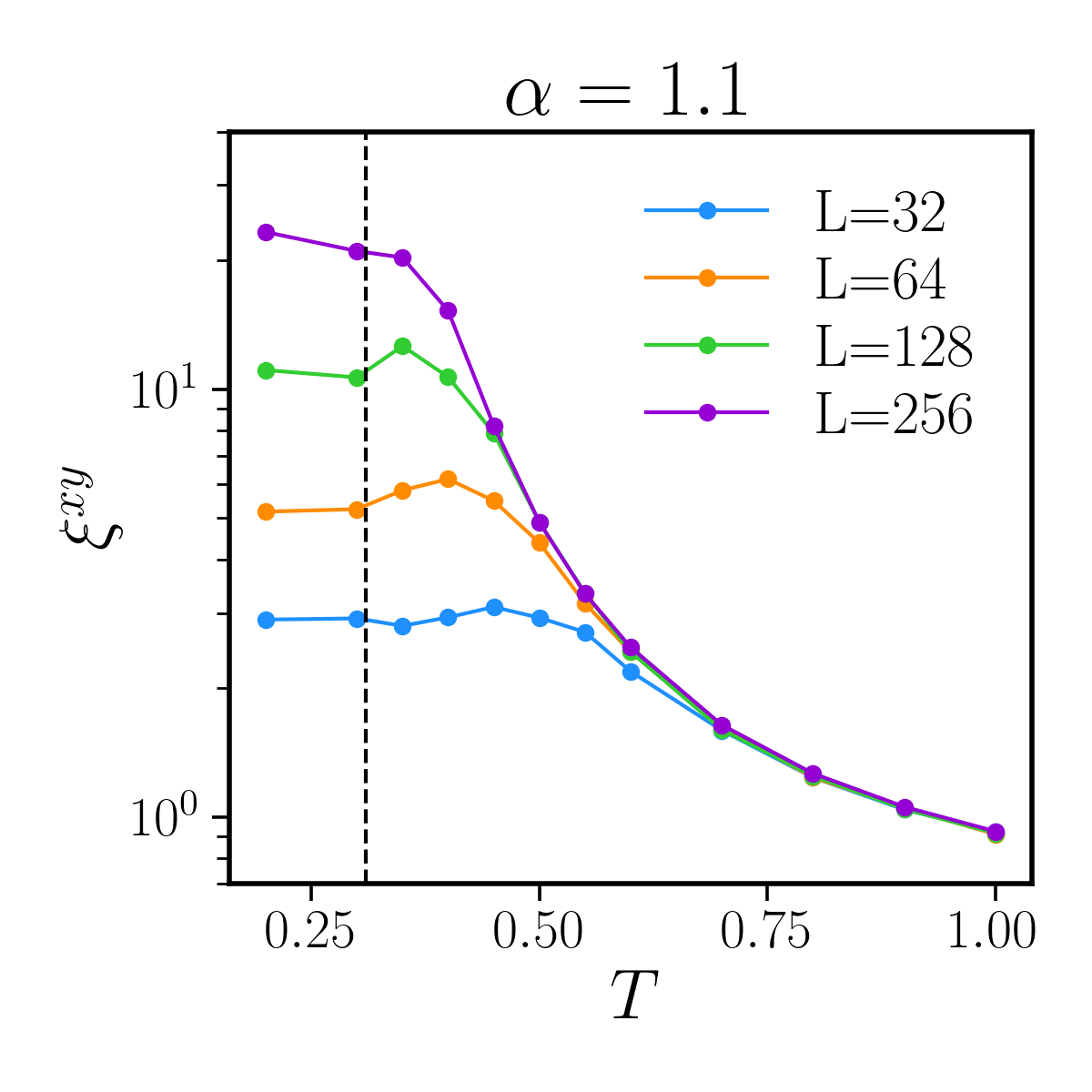}
    \includegraphics[width=0.24\textwidth]{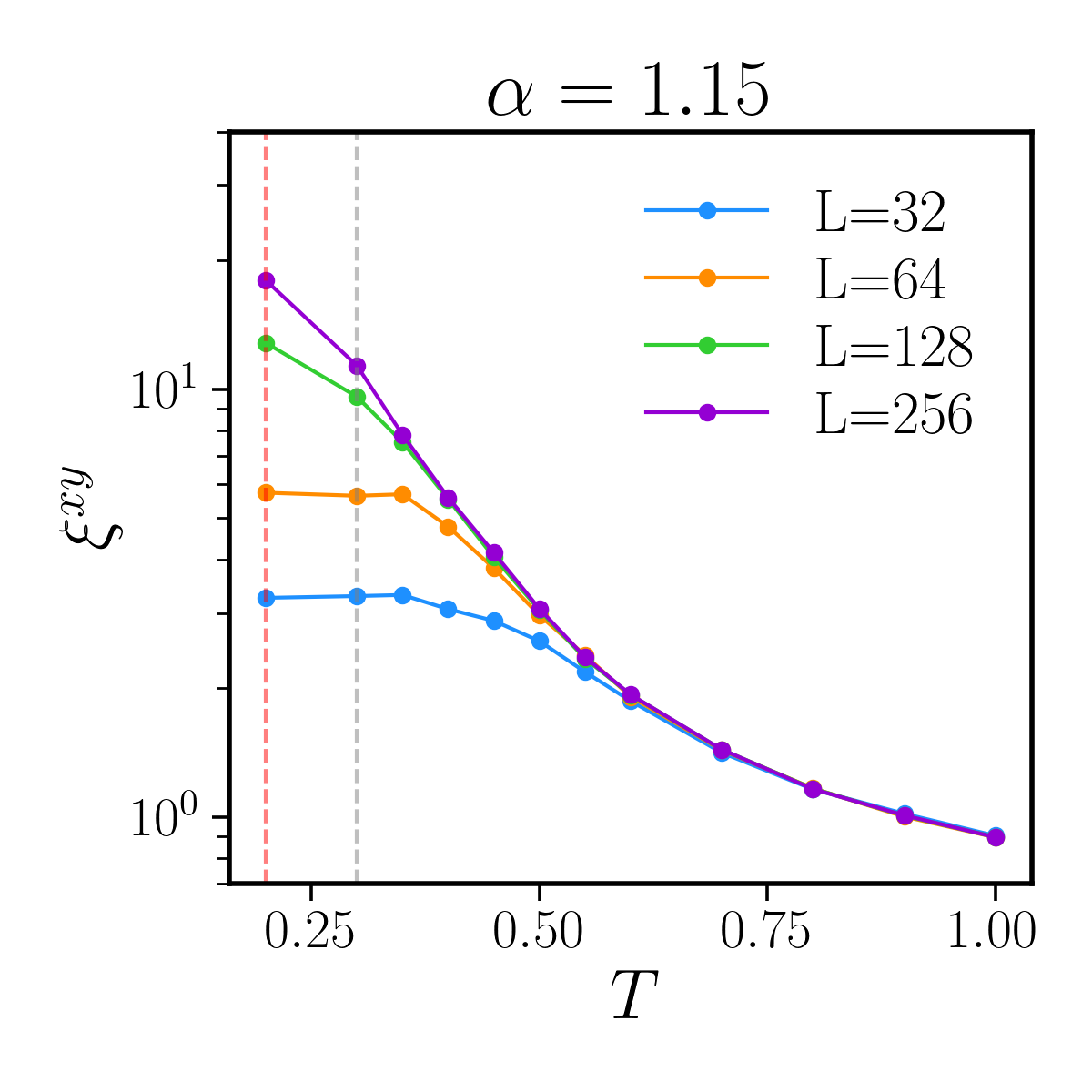}
    \caption{superconducting correlation length $\xi^{xy}$ as a function of temperature for $L=32,\,64,\,128,\,256$ at 
    various anisotropy parameter $\a=1,\,1.05\,1.1,\,1.15$. The vertical dashed black lines signal the critical 
    temperatures $T_{\text{BKT}}$, the dashed red line in the panel $\a=1.15$ at $T=0.2$ is the same temperature 
    indicated (also in red) in Fig.\,\ref{fig:Js_a1.15}; the grey dashed line at $T=0.3$ correspond to the maximum of $J_\text{s}$ for $L=16$. }
    \label{fig:xi_BKT}
\end{figure}

In order to investigate the role of superconducting phase fluctuations in this crossover filamentary state, 
we analyze the correlation length $\xi^{xy}$.
In Fig.\,\ref{fig:xi_BKT} we present $\xi^{xy}$ for $\alpha=1,\,1.05,\,1.1,\,1.15$ and $L=32,\,64,\,128,\,256$, as 
found by fitting the correlation function in Eq.\,\eqref{eq:C_bkt_XXZ}.
For $\a<1.15$, we find that $\xi^{xy}\sim L$, thus following the expected BKT scenario, thereby justifying the 
BKT analysis discussed above.
Note that the black dashed lines marks $T_\text{BKT}$ found from the crossing of $J_\text{s}$ 
(Fig.\,\ref{fig:FSC_Js}).
For $\a=1.15$, although $J_\text{s}$ vanishes at $L=128$, we can still observe BKT-like features of $\xi^{xy}$: 
the saturation value at low temperatures is in fact increasing with $L$, with some slowing down for $L=256$.
Again, we mark the temperature $T=0.2$ with a dashed red line: this temperature 
corresponds to the maximum of $J_\text{s}$ found at $L=16$ and to the change of slope in the decrease of 
$\overline{\< \sin\vp \>}$. The behaviour of  $\xi^{xy}(T)$ at this temperature seems to suggest the occurrence of an 
``avoided'' superconducting state, reminiscent of the findings in transport experiments~\cite{shi2014two,Leridon2020}.

\section{Three-dimensional phase diagram}
\label{sec:3d}

Starting from our detailed analysis of the two-dimensional model, we can make a comparison between the theoretical phase diagram Fig.~\ref{fig:phasediag_dirty} and the 
 experimental phase diagrams of the CO-SC competition of Fig.~\ref{fig:phasediag_exp}.
The most striking difference is that the CO temperature is strongly suppressed near the $p_{O(3)}$ point, both when compared with experiments and with the clean case. Also, FSC does not develop an evident foot in the CO region, although local SC regions are present.
To a large extent, both deficiencies can be ascribed to the 
low dimensionality of the model. To show this, we compute a three-dimensional phase diagram assuming an interlayer coupling both for the CO and the SC phases. 

As it is known from studies of the layered Heisenberg model,  the transition temperature can be estimated from the superconducting and CO correlation length ~\cite{Kastner1998} solving the following equations,
\begin{equation}
[\xi^{xy}(T_{SC}^{3D})]^2 J_{\perp}^{xy} = T_{SC}^{3D},    
\qquad
[\xi^z(T_{CO}^{3D})]^2 J_{\perp}^z = T_{CO}^{3D}    
\label{eq:T3D}
\end{equation}
 meaning that the interlayer energy associated with a correlated region of area $(\xi^{xy,z})^2$ is of the order of the critical temperature.

The superconducting and charge-ordered interlayer couplings, $J_{\perp}^{xy}$ and $J_{\perp}^z$ respectively, are not known. In view of the quasi 2D nature of cuprates we take  $J_{\perp}^z=0.1$ and  $J_{\perp}^{xy}=0.01$.
The much smaller value of the superconducting coupling with respect to the charge-ordered one is justified by the fact that CO is coupled by the long-range Coulomb interaction while SC is coupled by Josephson tunnelling through the insulating layers, and one expects a large difference between these two scales. For the rest, these parameters are rather arbitrary, but the qualitative form of the phase diagram is not expected to be sensitive to the precise value of the couplings.

\begin{figure}
    \centering
    \includegraphics[width=\linewidth]{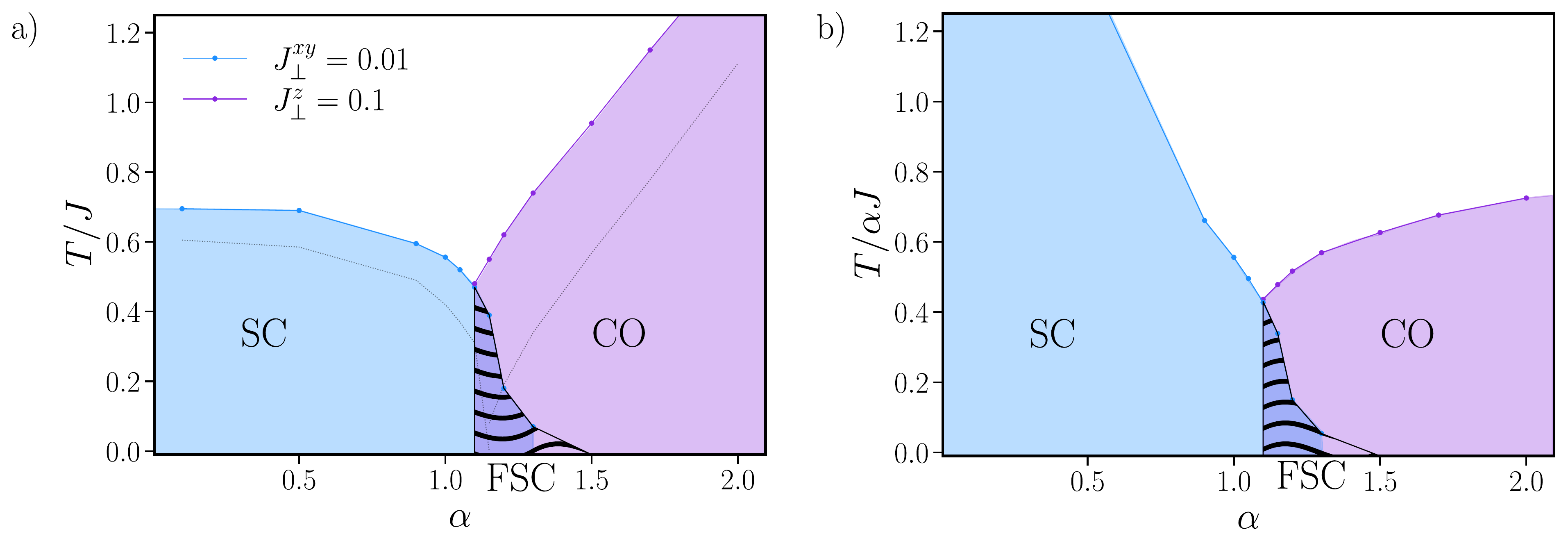}
    \caption{ (a) CO-driven  and (b) SC-driven 
    phase diagrams. Critical temperatures are normalized with respect to $J$ and $\alpha J$, respectively. The 3D critical temperatures are calculated from the correlation length according to Eq.\,\eqref{eq:T3D} with $J_\perp^{xy}=0.01$ and $J_\perp^{z}=0.1$. We specify that the FSC point in $\alpha=1.5$ is extrapolated from the vanishing of the in-plane superconducting component 
    (see purple curve in Fig.\,\ref{fig:Js_a1.15}b).
    For comparison, the grey lines in panel a) refer to the two-dimensional system.}
    \label{fig:phasediag_3D}
\end{figure}

Fig.\,\ref{fig:phasediag_3D}(a) shows the resulting phase diagram. The two-dimensional lines are showed with dotted grey lines for comparison. Panel (a) follows our previous convention of measuring energy and temperature in units 
of the superconducting scale $J$ so that the scale of the CO state changes with $\alpha$.  This corresponds to the CO-driven transition mentioned in the introduction. 
In panel b, by simply rescaling our energy units, we derive the phase diagram for the SC-driven transition. Here, the CO energy scale is, by definition, constant.  

We see that indeed the three-dimensional phase diagram bears a strong resemblance with the experimental phase diagrams for the SC-driven case (Fig.~\ref{fig:phasediag_exp}a,b) and CO-driven case (Fig. \ref{fig:phasediag_exp}c). 
Now the bicritical point is at a temperature of the order of the ordering temperature and the FSC foot extends more in the CO region.

\section{Conclusions}
\label{concl}

We used Monte-Carlo simulations to solve a statistical mechanics model  of a two-dimensional system presenting 
competition between SC and CO, both in the absence and in the presence of quenched disorder. We computed thermodynamic quantities, correlation functions, and thermodynamic phase diagrams. 

In a clean system, the competition mechanism generates metastability regions in the phase diagram, bounded by two spinodal 
lines and encompassing the first-order phase transition line.
As the temperature increases, the region 
of metastability shrinks to a single point, which coincides 
(within numerical accuracy) with the bicritical point, where the 
charge-ordered, superconducting and 
disordered phases meet. The first-order line separating the charge-ordered phase and the superconducting phase is rather steep
for low values of the barrier height ($B=0.2$), indicating 
that the two phases have similar entropy, as one can check using the 
Clausius-Clapeyron relation. We can thus make a comparison with the case 
of $^4$He \cite{zilsel1965pseudospin,liu1973hard}, where the almost vertical line separating the solid 
and superfluid phases led to the hypothesis that superfluidity was a 
low-entropy phase, as a crystal, fuelling explanations based on 
condensation in momentum space rather than in real space.

A closer inspection shows that the first-order line is not exactly 
vertical, and a re-entrance appears, thus showing that near $\alpha \gtrsim 1$ 
one can make a transition from the charge-ordered phase to 
the superconducting phase by increasing the temperature. This 
means that the superconducting phase has actually slightly higher 
entropy than the charge-ordered phase. A posteriori, this result 
is reasonable as the charge-ordered state has two gapped transverse 
modes while the superconducting state has one gapped mode and
one Goldstone mode. Thus, just considering low-laying excitations near 
$T = 0$, it is reasonable that the superconductor can have larger 
thermal fluctuations and entropy.
Interestingly, the re-entrant behaviour of the superconducting phase is 
also reminiscent of the phase diagram of $^4$He, in which 
a range of pressures is found where the solid $^4$He, if heated, transits 
to its superfluid state before becoming a simple liquid. 
In $^4$He, however, this happens in the high-temperature part of the 
phase diagram, while here we observe it at low temperatures.  
In fact, in the low-temperature region, the slope of our phase diagram and 
the one of $^4$He have opposite sign. 
We speculate that this qualitative difference is due to the fact that in our case the charge-ordered state has no 
Goldstone modes while in the 
case of $^4$He the crystal has sound (Goldstone) modes. 


The metastability regions and the first-order transition line disappear when quenched disorder is considered, giving rise instead to a phase-separated region where FSC appears.
Indeed, moving from the SC to the CO regime 
we find the gradual disappearance of the two-dimensional superconducting phase towards a 
polycrystalline charge-ordered phase with the tuning of the anisotropy parameter $\a$. As the BKT signatures disappear, 
one-dimensional-like superconducting patterns still survive inside the charge-ordered phase.

In Refs.\,\cite{attanasi2009competition,Leridon2020,caprara2020doping} it was 
already proposed that disorder may have a peculiar effect 
in the coexistence region discussed above, turning the 
metastable superconducting state into a stable state, where 
FSC is topologically protected at 
the boundaries between different charge-ordered domains, in agreement 
with the tentative phase diagram proposed for cuprates 
in Ref.\,\cite{caprara2020doping}. Such a phase diagram 
was purely based on the peculiarities of the resistance curves as 
a function of the temperature, with varying magnetic field and doping, and showed that 
SC can develop at low temperatures even when at 
high temperature the system is well inside the charge-ordered region 
of the phase diagram. In this work, we provided a solid background 
to the above scenario, showing that within our minimal model for the 
competition between CO and SC, 
Eq.\,(\ref{eq:XXZ_Htot}), a random magnetic field has exactly the effect of
promoting the fragmentation of the charge-ordered state into domains 
exhibiting the two different realizations of CO. In the 
domain wall they frustrate each other resulting in the stabilization of the superconducting state. Once 
this FSC is suppressed (by increasing the 
temperature and/or the non-thermal parameter $\alpha$), only 
polycrystalline CO remains. The polycrystalline CO  is characterized by large puddles with different realizations of CO, and it resembles the complex landscape of charge-ordered domains experimentally observed in cuprates \cite{campi2015inhomogeneity}. 

The filamentary superconductivity foot we find in the two-dimensional phase diagram (Fig.~\ref{fig:phasediag_dirty}) is relatively small compared with experiments (Fig.~\ref{fig:phasediag_exp}). Also, the CO temperature is strongly suppressed close to the 
$O(3)$ point in the presence of quenched disorder. Both features are cured considering an interlayer coupling, yielding a phase diagram nicely resembling experiments, both in the CO- and SC-driven case. 


Very near the $O(3)$ point in the dirty case, we find that the superfluid stiffness has a non-monotonous behaviour as a function of $T$ with a maximum at intermediate temperatures (Fig.~\ref{fig:FSC_Js}c). It would be very interesting to observe experimentally this effect as it would be a signature of entropically favoured superconductivity.   

In our model, up and down pseudospins encode  only two possible realizations of CO corresponding to a checkerboard pattern in a bipartite lattice. This is a simplification of cuprates  where,  for example, non-magnetic charge stripes with periodicity four have four CO variants for each orientation, yielding 16 possible ``colours"  of CO patterns. Still, our simplified ``two-colour" model captures many subtleties of the phase diagram.

The presence of some one-dimensional-like superconducting patterns persisting on the CO side of the phase diagram can 
indeed have
striking effect on the macroscopic observable, such as specific heat \cite{Kacmarcik2018} or spin susceptibility 
\cite{zhou2017spin}, and particularly on transport measurements \cite{shi2014two, shi2020vortex, caprara2020doping}.


\section*{Acknowledgements}
We acknowledge stimulating discussions with M. Grilli, B. Leridon, F. Ricci-Tersenghi.


\paragraph{Funding information}
We acknowledge financial support from the University of Rome Sapienza, 
under the projects Ateneo 2020 (RM120172A8CC7CC7), Ateneo 2021 (RM12117 A4A7FD11B), 
Ateneo 2022 (RM12218162CF9D05), from the Italian Ministero
dell’U\-ni\-ver\-si\-t\`a e della Ricerca, under the Project
PRIN 2017Z8TS5B,  PRIN 20207ZXT4Z, from PNRR MUR project PE0000023-NQSTI, and from CNR-CONICET project ``New materials for superconducting electronics".

\begin{appendix}

\section{Superfluid stiffness}
\label{app:Js}

\begin{figure}[!h]
    \centering
    \includegraphics[width=0.8\linewidth]{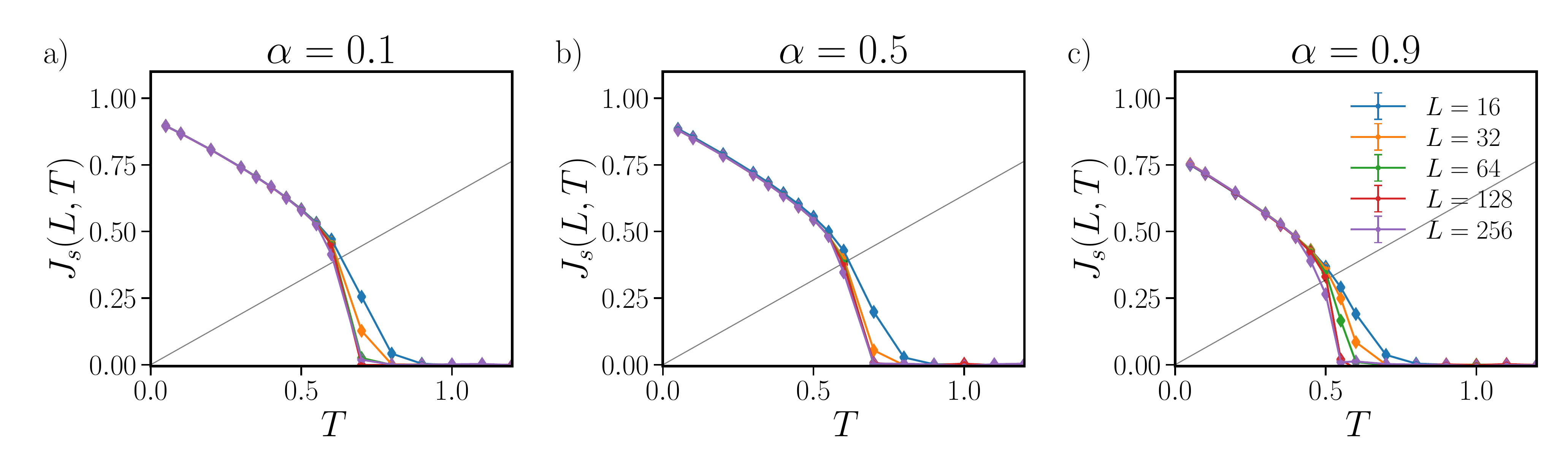}\\
    \includegraphics[width=0.8\linewidth]{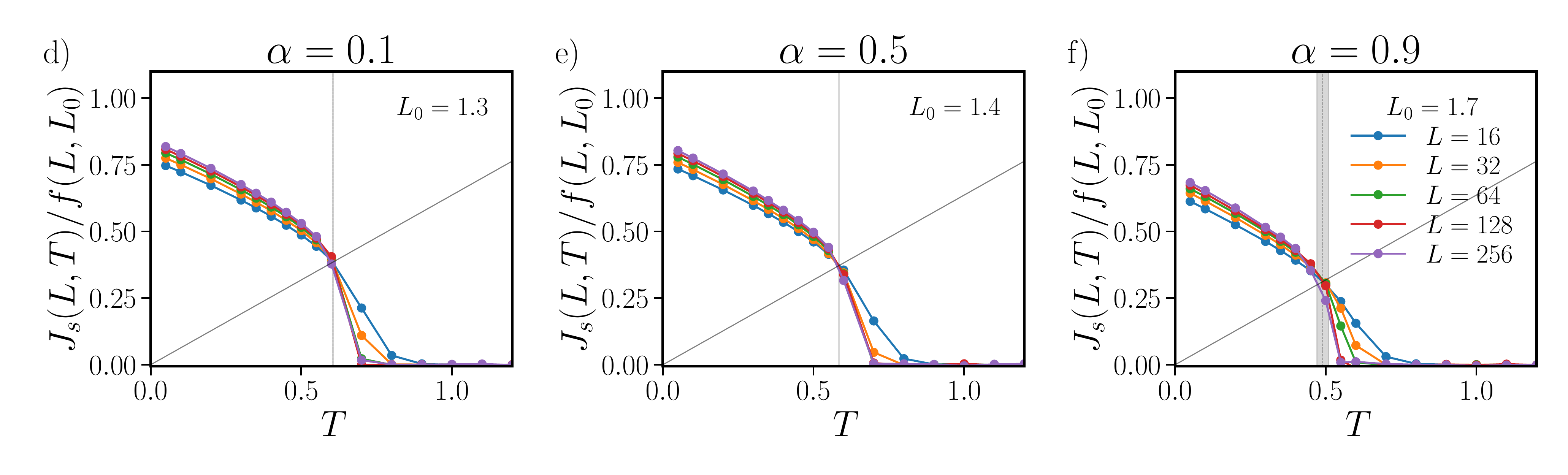}
    \caption{(a) Superfluid stiffness and (b) rescaled superfluid stiffness ($f(L,L_0)=1+[2\ln(L/L_0)]^{-1}$, see 
    Eq.\,\ref{eq:XXZ_scalingBKT}) for $\a=0.1,\,0.5,\,0.9$. Errorbars in panel b refers to the standard deviation computed on different indipendent disorder 
    realizations. The vertical grey lines indicates $T_\text{BKT}$, black lines are the universal critical line $2 
    T/\pi$.}
    \label{fig:app1}
\end{figure}

For the sake of completeness, we show the superfluid stiffness and their BKT critical jump for all the values of $\alpha$ 
considered to construct our phase diagram (Fig.\,\ref{fig:phasediag_dirty}).
In Fig.\,\ref{fig:app1} it is possible to observe the validity of Harris criterion when addressing the BKT transition for 
$\a<1$, where the disorder leaves $J_s$ almost unaffected. 
The only appreciable effect relies in the suppression of both the saturating value of $J_s$ for $T\ra0$ (see panels a, b 
and c), which is lowered to $0.75$ for $\a=0.9$ while the critical temperature is only very slightly decreased. 
This can be appreciated looking at panels d, e, and f, where we show the relative crossing points with the universal 
critical line $2T/\pi$, indicating with a vertical line the corresponding $T_\text{BKT}$.
A first consequence of the random field is indeed visible in the smearing of this crossing at $\a=0.9$, highlighted in 
grey.

\begin{figure}[!h]
    \centering
    \includegraphics[width=0.9\linewidth]{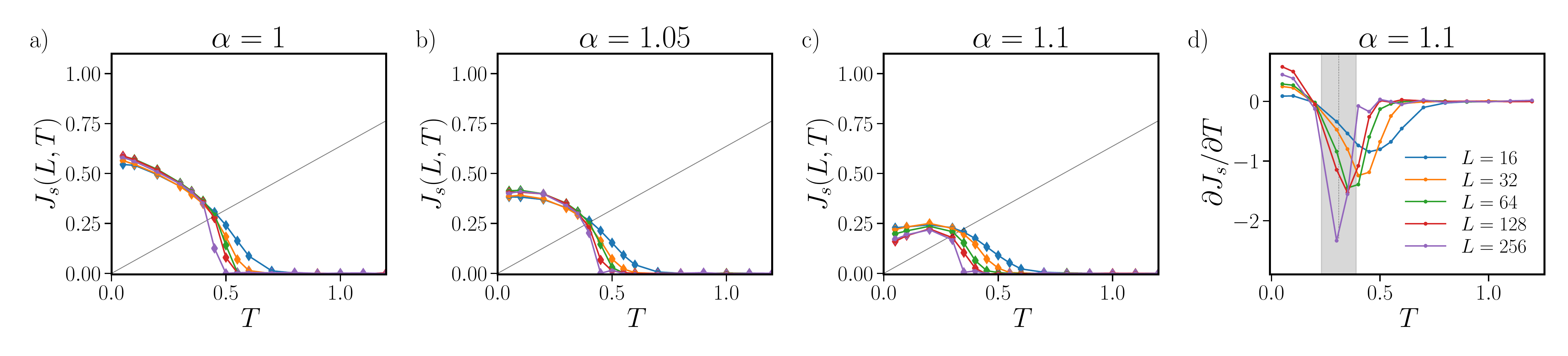}
    \caption{Unscaled superfluid stiffness with errorbars for (a) $\a=1$, (b) $\a=1.05$, (c) $\a=1.1$ and (d) its first 
    derivative with respect to the temperature $\partial J_s/\partial T$. The errorbars in panels a, b and c are calculated from the standard deviation of independent disorder configurations.  In panel d the vertical line and the grey 
    shaded area indicates $T_\text{BKT}$ with its error, extracted using the BKT scaling law and showed in 
    Fig.\,\ref{fig:FSC_Js}.}
    \label{fig:app3}
\end{figure}

In panels a, b and c of Fig.\,\ref{fig:app3} instead we present the superfluid stiffness in the filamentary region of 
our phase diagram, namely $\a=1,\,1.05,\,1.1$.
The suppression of $J_s$ caused by the presence of the correlated disorder emerging is much more visible here.
In particular, we highlight the fact that, while for $\a=1,\,1.05$ the scaling law still produces efficient results (see 
panels a and b of Figs.\,\ref{fig:app1} and \ref{fig:FSC_Js}), this does not seem to be the case for $\a=1.1$ ((see 
panels a and b of Figs.\,\ref{fig:app3} and \ref{fig:FSC_Js}).
However, the extrapolated $T_\text{BKT}$ is consistent with the minimum found for its derivative 
$\partial J_s/\partial T$ \cite{lee2005helicity, lee2005monte}.
This is shown in panel d where the vertical line is at $T_\text{BKT}$ and the grey area highlights the error considered.

\end{appendix}
\vspace{1cm}



\bibliographystyle{SciPost_bibstyle}
\bibliography{giulia, library}

\nolinenumbers

\end{document}